\documentclass[lettersize,journal]{IEEEtran}
\IEEEoverridecommandlockouts
\usepackage{cite}
\usepackage{color}
\usepackage{amsthm}
\usepackage{amsmath,amssymb,amsfonts}
\usepackage{bbm}
\usepackage{dsfont}
\usepackage{algorithmic}
\usepackage{graphicx}
\usepackage{textcomp}
\usepackage{xcolor}
\usepackage{amsthm,amsmath,amssymb}
\usepackage{steinmetz}
\usepackage{textcomp}
\usepackage{mathrsfs}
\usepackage{amsmath,bm}
\usepackage{algorithm} 
\usepackage{algorithmic}
\usepackage{mathtools}
\usepackage{epsf}
\usepackage{cleveref}
\usepackage{verbatim}
\usepackage{subfigure}
\usepackage{tikz}
\usepackage[justification=centering]{caption}

\newtheorem{lemma}{Lemma}
\newtheorem{prop}{Proposition}

\allowdisplaybreaks
\def\BibTeX{{\rm B\kern-.05em{\sc i\kern-.025em b}\kern-.08em
    T\kern-.1667em\lower.7ex\hbox{E}\kern-.125emX}}
\begin{document}

\title{ 
     Over-the-Air Modulation for RIS-assisted Symbiotic Radios: Design, Analysis, and Optimization
}
\author{
\IEEEauthorblockN{Hu~Zhou,~Ying-Chang Liang,~\IEEEmembership{Fellow,~IEEE},~and~Chau~Yuen,~\IEEEmembership{Fellow,~IEEE}}
\\
\thanks{This work has been submitted to the IEEE for possible publication. Copyright may be transferred without notice, after which this version may no longer be accessible.}
\thanks{
Part of this work was presented in IEEE Globecom 2023~\cite{zhou2023}. H.~Zhou is with the National Key Laboratory of Wireless Communications, and the Center for Intelligent Networking and Communications (CINC), University of Electronic Science and Technology of China (UESTC), Chengdu 611731, China (e-mail: {huzhou@std.uestc.edu.cn}). Y.-C. Liang is with the Center for Intelligent Networking and Communications (CINC), University of Electronic Science and Technology of China (UESTC), Chengdu 611731, China (e-mail: {liangyc@ieee.org}). C. Yuen is with the School of Electrical and Electronics Engineering, Nanyang Technological University, 639798 Singapore (e-mail: chau.yuen@ntu.edu.sg).}
}
% \author{
% \IEEEauthorblockN{Hu~Zhou,~and~Ying-Chang Liang,~\IEEEmembership{Fellow,~IEEE}
% }
% \\
% \thanks{
% Part of this work has been submitted to IEEE Globecom 2023~\cite{zhou2023}.}
% \\
% \thanks{H.~Zhou and Y.-C. Liang are with University of Electronic Science and Technology of China (UESTC), Chengdu 611731, China (e-mails: {huzhou@std.uestc.edu.cn;~liangyc@ieee.org}).}
% }
\maketitle

\begin{abstract}
In reconfigurable intelligent surface (RIS)-assisted symbiotic radio (SR), an RIS is exploited to assist the primary system and to simultaneously operate as a secondary transmitter by modulating its own information over the incident primary signal from the air. Such an operation is called over-the-air modulation. The existing modulation schemes such as on-off keying and binary phase-shift keying suffer from two problems for joint detection of the primary and secondary signals in RIS-assisted SR, i.e., one is the detection ambiguity problem when the direct link is blocked, and the other is the bit error rate (BER) error-floor problem when the direct link is weak. To address the two problems, we propose a novel modulation scheme by dividing the phase-shift matrix into two parts: one is the assistance beamforming matrix for assisting the primary system and the other is the transmission beamforming matrix for delivering the secondary signal. To optimize the assistance and transmission beamforming matrices, we first introduce an assistance factor that describes the performance requirement of the primary system and then formulate a problem to minimize the BER of the secondary system, while guaranteeing the BER requirement of the primary system controlled by the assistance factor. To solve this non-convex problem, we resort to the successive convex approximation technique to obtain a suboptimal solution. Furthermore, to draw more insights, we propose a low-complexity assistance-transmission beamforming structure by borrowing the idea from the classical maximum ratio transmission and zero forcing techniques. Finally, simulation results reveal an interesting tradeoff between the BER performance of the primary and secondary systems by adjusting the assistance factor.
\end{abstract}

\vspace{-0.15cm}
\section{Introduction}
Recent years have witnessed the rapid development of the Internet of Things (IoT), which plays a vital role in shaping our daily lives, ranging from the classical applications, e.g., smart home and smart city, to the newly emerged applications, such as augmented/virtual reality, autonomous driving, and holographic communications~\cite{nguyen20216g}. As predicted by Cisco, the sixth-generation (6G) wireless networks are envisioned to support billions of IoT devices by 2030, which poses significant challenges in terms of the scarce spectrum and energy resources~\cite{dang2020should}. 

Recently, symbiotic radio (SR), has been proposed as a spectrum- and energy-efficient technology to address the above challenges associated with IoT~\cite{long2019symbiotic,liang2020symbiotic}. In SR, a passive secondary system, e.g., IoT transmission, is parasitic in an active primary system, e.g., cellular communication. Particularly, the backscatter device (BD) acts as an IoT device to transmit its sensed environmental information by backscattering the incident primary signal without generating an RF carrier, which is therefore spectrum- and energy-efficient~\cite{wang2024multi,dai2022rate}.
However, in general, only a single or limited number of antennas are deployed at the BD, which limits the performance of the backscatter communication in SR.
Fortunately, the emergence of reconfigurable intelligent surfaces (RISs) offers a solution to solve this problem.
Specifically, RIS is made up of multiple electromagnetic reflecting elements, each of which is able to reflect the incident signal with an adjustable phase shift~\cite{wu2019towards,huang2019reconfigurable,liang2019large}. Considering the similar reflection principle between the RIS and backscatter communication, recent research proposes to introduce RIS into SR as a BD to further enhance the performance of SR by exploiting its multiple reflecting elements~\cite{lei2021reconfigurable,liang2022backscatter}.

In RIS-assisted SR, the RIS is used for not only assisting the primary system by providing an additional reflecting link but also for acting as a secondary transmitter (ST) to deliver its information to the secondary user. To be specific, the RIS can modulate its own information bits over the incident primary signal from the air. Therefore, such an operation is called over-the-air modulation that shares the spectrum and energy resources with the primary system. 
Notably, one important feature of over-the-air modulation is that the reflecting signal out of the RIS is the multiplication of the primary and secondary signals. Thus, the reflecting link contains both the information of the primary and secondary signals. As a reward for utilizing the spectrum and energy of the primary system, the secondary system via the RIS can assist the primary system by passive beamforming~\cite{zhang2024channel,chen2023transmission}. Therefore, the primary and secondary systems in RIS-assisted SR form a mutualistic spectrum- and energy-sharing relationship.
Moreover, depending on whether the primary user (PU) and secondary user (SU) are spatially co-located or separated, RIS-assisted SR can be classified into integrated model and detached model. For the integrated model, the PU and the SU are integrated as a cooperative receiver, which is used to jointly decode the primary and secondary signals. While for the detached model, the PU and the SU are separated to decode their individual signals, respectively.

In the literature, many research efforts have been devoted to over-the-air modulation schemes for both the integrated and detached models of RIS-assisted SR~\cite{yan2020passive,hua2021novel,hua2021uav,zhang2021reconfigurable,zhou2022cooperative,guo2020reflecting,wu2021reconfigurable,li2022ris,lin2020reconfigurable,lin2021reconfigurable,li2022reconfigurable}, such as on/off keying (OOK) modulation~\cite{yan2020passive,hua2021novel,hua2021uav}, binary phase shift keying (BPSK) modulation~\cite{zhang2021reconfigurable,zhou2022cooperative}, reflecting modulation~\cite{guo2020reflecting}, spatial modulation~\cite{wu2021reconfigurable,li2022ris,lin2020reconfigurable,lin2021reconfigurable}, and number modulation~\cite{li2022reconfigurable}. 
For example, considering the integrated model, OOK modulation is proposed by carrying the RIS information via the on/off states of reflecting elements~\cite{yan2020passive,hua2021novel,hua2021uav}. Then, considering that turning off part of the reflecting elements would reduce the passive beamforming gain, the authors of~\cite{lin2020reconfigurable,lin2021reconfigurable} propose to fix the number of closed reflecting elements or turning on all the reflecting elements with different phase shifts. As for BPSK modulation, all the reflecting elements are turned on and the RIS transmits information by switching two types of phase shifts with an opposite phase~\cite{zhang2021reconfigurable,zhou2022cooperative}. As for reflecting modulation, a general framework is proposed in~\cite{guo2020reflecting}, where the signal mapping, shaping, and phase shifts are jointly designed to support RIS information transmission. To further enhance the spectral efficiency, spatial modulation is proposed in~\cite{wu2021reconfigurable} that transmits information by jointly exploiting the receive-antenna index and RIS phase shifts. For the number modulation, the RIS is also partitioned into two sub-surfaces with orthogonal phase while the RIS information is carried by exploiting the number of reflecting elements of the first/second sub-surface~\cite{li2022reconfigurable}. 
Moreover, considering the detached model, the authors propose a spatial modulation scheme where the RIS information is embedded in the receive-antenna index of the secondary user~\cite{li2022ris}.
Based on the above modulation schemes, joint active and passive beamforming design for RIS-assisted SR has been extensively studied~\cite{wang2021intelligent,zhang2021reconfigurable,hu2020reconfigurable,hua2021novel,wu2023ris,ye2020joint}, e.g.,  power minimization, BER minimization, and achievable rate maximization. 

Despite the research progress for RIS-assisted SR, over-the-air modulation 
has yet to be well studied. In this paper, we will unveil two fundamental challenges faced by over-the-air modulation. The first one is the detection ambiguity problem in the absence of the direct link. This is because when the direct link is blocked, the receiver only receives the multiplicative signal of the primary and secondary ones from the reflecting link. In this case, it is impossible for the receiver to jointly detect the primary and secondary signals due to the ambiguity caused by signal multiplication.
The second one is the BER error floor problem when the direct link is weak. In this case, if we increase the number of reflecting elements (i.e., $N$) such that the reflecting link is stronger than the direct link, the BER of the primary and secondary signals will converge to a fixed value irrespective of $N$. This implies that no matter how large $N$ is, it does not help enhance the BER performance of RIS-assisted SR. Moreover, most existing studies focus on the over-air-modulation for the integrated model of RIS-assisted SR, which, however, is a special case of the detached model. A more general scenario is that the primary and secondary users are separated and the RIS modulation should take into account the beamforming direction towards the primary and secondary users. Therefore, it is necessary to carry out the study on the general detached model.

To address the above challenges, in this paper, we propose a novel over-the-air modulation scheme for the detached model of RIS-assisted SR where the primary and secondary users are spatially separated. The proposed scheme mathematically transforms the RIS phase-shift matrix into the summation of two matrices, where one is the assistance beamforming matrix used for assisting the primary system and the other is the transmission beamforming matrix used for delivering the secondary signal. The advantages of the proposed scheme are three-fold. First, with the proposed scheme, the ambiguity problem can be easily addressed since the reflecting link via the assistance beamforming matrix can be viewed as an equivalent direct link when the direct link is blocked. Second, the BER error floor problem can also be addressed since the assistance beamforming matrix can enhance the equivalent direct link when the direct link is weak.
Third, the proposed scheme offers us more flexibility to balance the performance of the primary and secondary systems by optimizing the assistance and transmission beamforming matrices. In a nutshell, the main contributions of this paper are summarized as follows.

\begin{itemize}
    \item We identify two fundamental challenges for over-the-air modulation in RIS-assisted SR, i.e., one is the detection ambiguity problem when the direct link is blocked and the other is the BER error floor problem when the number of reflecting elements is large. 
    \item To address the two challenges, we propose a novel over-the-air modulation scheme for RIS-assisted SR, which divides the phase-shift matrix into two beamforming matrices, i.e., one is the assistance beamforming matrix used to assist the primary system and the other is the transmission beamforming matrix used to transmit the secondary signal.
    \item  To optimize the assistance and transmission beamforming matrices, we first theoretically analyze the BER performance of the primary and secondary systems when their receivers are separated. Then, we formulate a problem to minimize the BER of the secondary system, under the constraint that the BER performance of the primary system is higher than its minimum requirement.
    \item To gain more insights into RIS-assisted SR, we propose a low-complexity assistance-transmission beamforming structure by borrowing the idea from classic maximum-ratio-transmission (MRT)~\cite{lo1999maximum} and zero-forcing (ZF) techniques~\cite{song2013prior}. Notably, under this beamforming structure, the secondary system can achieve its transmission without causing any impact on the primary system, while the primary system can still obtain assistance from RIS.
    \item Finally, simulation results show that our proposed scheme can strike a flexible balance between the BER performance of the primary and secondary systems by adjusting the assistance and transmission beamforming matrices.
\end{itemize}

The rest of this paper is organized as follows. Section \ref{sec-system-model} introduces the system model and the proposed over-the-air modulation scheme. Section \ref{sec: performance-analysis} analyzes the BER performance and reveals the fundamental problems of the existing modulation schemes.
Section \ref{sec:problem-formulation} presents the problem formulation. Then, Section \ref{sec: proposed-algorithm} proposes algorithms to solve the problem. Section \ref{sec:theorectical-analysis} proposes a low-complexity assistance-transmission beamforming structure. Section \ref{sec-simulation-results} provides simulations to evaluate the proposed scheme and algorithms. Finally, Section \ref{sec-conslusion} concludes this paper.

\emph{Notations}: The scalar, vector, and matrix are denoted by the lowercase, boldface lowercase, and boldface uppercase letters, respectively.  $\mathbb{C}^{x \times y} $ represents the space of $ x \times y $ complex-valued matrices.
$\mathcal{CN}(\mu,\sigma^{2})$ denotes complex Gaussian distribution with mean $\mu$ and variance $\sigma^{2}$. 
$\jmath$ denotes the imaginary unit. $\angle x$, $x^{H}$, and $\Re\{x\}$ denote the phase, conjugate transpose, and the real part of number $x$, respectively. $\mathcal{Q}(t)\triangleq\int_{t}^{\infty}\frac{1}{\sqrt{2\pi}}e^{-\frac{1}{2}\eta^{2}}d\eta$ is the complementary distribution function of the standard Gaussian. $\mathrm{diag}(\bm{a})$ denotes a diagonal matrix with its diagonal elements being the vector $\bm{a}$. $[\bm{X}]_{m,n}$ denotes the $(m,n)$-th element of matrix $\bm{X}$. $\mathrm{Tr}(\bm{X})$ and $\mathrm{rank}(\bm{X})$ denote the trace and the rank of matrix $\bm{X}$, respectively. $\mathrm{vec}(\cdot)$ denotes the vectorization operation and $\otimes$ denotes the Kronecker product. $\mathbf{Proj}_{\mathcal{S}} \bm{x}$ denotes the projection of $\bm{x}$ onto the set of $\mathcal{S}$.

\section{System Model}
\label{sec-system-model}
\begin{figure}[t]  
	\centering  
	% \captionsetup{font={scriptsize}}
	% \setlength{\belowcaptionskip}{-0.1cm}
	\includegraphics[width=3in]{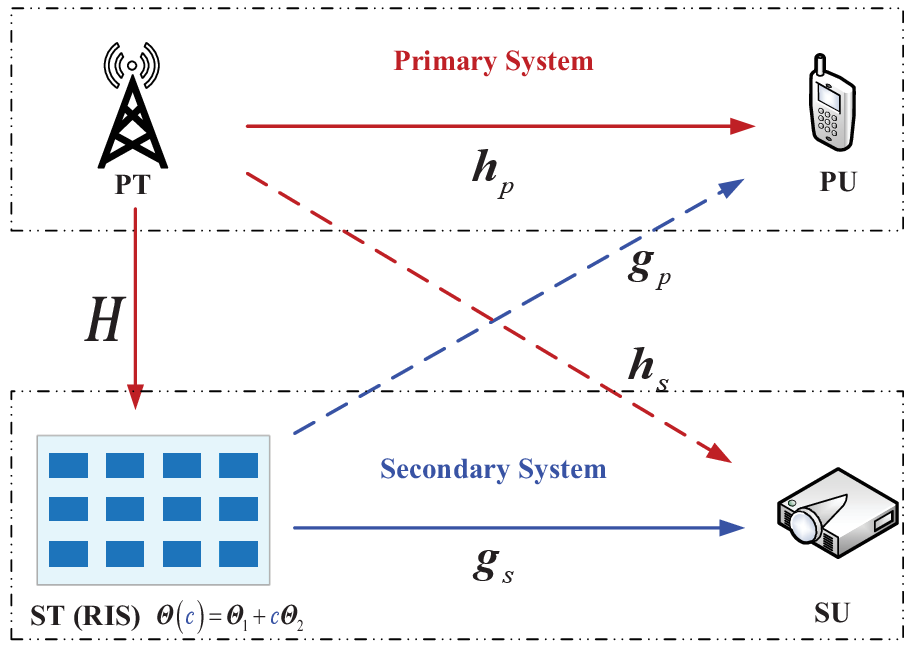}
	\caption{System model of RIS-assisted symbiotic radios.} 
	\label{fig:system-model}  
\end{figure}
As shown in Fig. \ref{fig:system-model}, we consider the RIS-assisted SR system. The $M$-antenna primary transmitter (PT) communicates with the primary user (PU) by using the active radio technology, e.g., cellular communications. By sharing the spectrum and energy resource of the primary system, the secondary system makes use of passive radio technology to transmit information from the secondary transmitter (ST) to the secondary user (SU), e.g., IoT communications. Particularly, the  RIS, equipped with $N$ reflecting elements, plays the role of ST, which simultaneously assists the primary system and transmits the secondary signal.
In the $l$-th symbol period, the primary and secondary signals are denoted by $s(l)\!\in\!\mathcal{A}_{s}$ and $c(l)\!\in\!\mathcal{A}_{c}$, where $\mathcal{A}_{s}$ and $\mathcal{A}_{c}$ are their constellation sets, respectively.
In what follows, we will introduce the proposed over-air-modulation scheme, the signal model, and the corresponding receiver design, respectively.

\subsection{Over-the-Air Modulation}
It is known that RIS does not generate RF signals by itself but passively reflects the incident signals. Therefore, when RIS acts as a transmitter, it modulates its information bits over the incident RF signal from the air. Such a scheme is also called over-the-air modulation.
In this paper, we adopt the modulation scheme proposed in our earlier work~\cite{zhou2023modulation}, which divides the RIS phase-shift matrix into two parts: one is named the assistance beamforming matrix used to assist the primary system, while the other is named the transmission beamforming matrix used to transmit the secondary signal. By doing so, the mapping rule between the secondary signal $c(l)$ and the phase-shift matrix $\bm{\Theta}(c(l))\in\mathbb{C}^{N\times N}$ can be modeled as
\begin{align}\label{eq: phase-shifts-mapping}
    \bm{\Theta}(c(l))=\bm{\Theta}_{1}+c(l)\bm{\Theta}_{2}.
\end{align}

In \eqref{eq: phase-shifts-mapping}, $\bm{\Theta}_{1}\!=\!\mathrm{diag}(\theta_{1,1},\cdots,\theta_{1,N})\!\in\!\mathbb{C}^{N\times N}$ is the assistance beamforming matrix, and $c(l)\bm{\Theta}_{2}\!=\!c(l)\mathrm{diag}(\theta_{2,1},\cdots,\theta_{2,N})\in\mathbb{C}^{N\times N}$, where $\bm{\Theta}_{2}$ the transmission beamforming matrix of signal $c(l)$. During the channel coherence time, both $\bm{\Theta}_{1}$ and $\bm{\Theta}_{2}$ are fixed while the secondary signal $c(l)$ is changed according to the secondary signal to be transmitted. In practice, such a modulation scheme can be realized as follows.
Take $\mathcal{A}_{c}\!=\!\{1,-1\}$ as an example. When RIS transmits symbol $c(l)=1$, the RIS sets its phase-shift matrix as $\bm{\Theta}_{1}+\bm{\Theta}_{2}$. While RIS transmits symbol $c(l)=-1$, the phase-shift matrix is set to $\bm{\Theta}_{1}-\bm{\Theta}_{2}$. By switching between the two phase-shift matrices, i.e., $\bm{\Theta}_{1}+\bm{\Theta}_{2}$ and $\bm{\Theta}_{1}-\bm{\Theta}_{2}$, the secondary signal can be delivered. Note that when $\bm{\Theta}_{1}=\bm{0}$, the proposed scheme degrades to the existing scheme $\bm{\Theta}(c(l))=c(l)\bm{\Theta}_{2}$. The two fundamental problems associated with the existing scheme, i.e., detection ambiguity and error floor problems, will be detailed in Sec. \ref{sec:performance}.
\subsection{Received Signal Model}
The block flat-fading channels are assumed in this paper. In practice, the channel state information (CSI) can be estimated via training-based techniques~\cite{wu2019towards}. 
The channel coefficients from the PT to the PU and from the PT to the SU are denoted by $\bm{h}_{p}\in\mathbb{C}^{M\times 1}$ and $\bm{h}_{s}\in\mathbb{C}^{M\times 1}$, respectively. The channel coefficients
from the PT to the ST (RIS), from the ST (RIS) to the PU, and from the ST to the SU are denoted by $\bm{H}\in\mathbb{C}^{N\times M}$, $\bm{g}_{p}\in\mathbb{C}^{N\times 1}$, and $\bm{g}_{s}\in\mathbb{C}^{N\times 1}$, respectively. Denote by $\bm{w}\in\mathbb{C}^{M\times1}$ the linear beamforming vector at the PT. Then, the received signals at the PU and the SU are given by
\begin{align} \label{eq: received-signal-PU}
    y_{p}(l)&=\underbrace{\bm{h}_{p}^{H}\bm{w}s(l)}_{\text{Direct link}}+\underbrace{\bm{g}_{p}^{H}\bm{\Theta}_{1}\bm{H}\bm{w}s(l)}_{\text{Reflecting link for assistance}} \nonumber \\
    & \quad +\underbrace{\bm{g}_{p}^{H}\bm{\Theta}_{2}\bm{H}\bm{w}c(l)s(l)}_{\text{Reflecting link for transmission}}+z_{p}(l),
\end{align}
and
\begin{align} \label{eq: received-signal-SU}
    y_{s}(l)=&\bm{h}_{s}^{H}\bm{w}s(l)+\bm{g}_{s}^{H}\bm{\Theta}_{1}\bm{H}\bm{w}s(l)
    \nonumber \\
    & +\bm{g}_{s}^{H}\bm{\Theta}_{2}\bm{H}\bm{w}c(l)s(l)+z_{s}(l),
\end{align}
% \begin{align} \label{eq: received-signal-SU}
%     y_{s}(l)=&\underbrace{\sqrt{P_t}h_{s}s(l)}_{\text{Direct link}}+\underbrace{\sqrt{P_t}\bm{g}_{s}^{H}\bm{\Theta}_{1}\bm{h}s(l)}_{\text{Reflected link for assistance}}\nonumber\\
%     &+\underbrace{\sqrt{P_t}\bm{g}_{s}^{H}\bm{\Theta}_{2}\bm{h}c(l)s(l)}_{\text{Reflected link for transmission}}+z_{s}(l),
% \end{align}
respectively; $z_{p}(l)\sim \mathcal{CN}(0,\sigma_{p}^{2})$ and $z_{s}(l)\sim \mathcal{CN}(0,\sigma_{s}^{2})$ denote the complex Gaussian noises at the PU and the SU with zero mean and variances $\sigma_{p}^{2}$, $\sigma_{s}^{2}$, respectively.  
By adopting our proposed modulation scheme, the signal model is different from those in~\cite{zhang2021reconfigurable,zhou2022cooperative,hua2021novel}.
Specifically, the received signals in \eqref{eq: received-signal-PU} and \eqref{eq: received-signal-SU} consists of three parts: the primary signal $s(l)$ from the direct link, the primary signal $s(l)$ from the reflecting link via $\bm{\Theta}_{1}$, and the product of the primary and secondary signals $s(l)c(l)$ from the reflecting link via $\bm{\Theta}_{2}$.

\subsection{Receiver Design}
As can be seen from \eqref{eq: received-signal-PU} and \eqref{eq: received-signal-SU}, the primary and secondary signals are coupled in a both additive and multiplicative manner at the PU and the SU, which makes it quite challenging to decode them. Considering that in RIS-assisted SR, there exists the collaboration between the primary and secondary systems~\cite{liang2022backscatter,zhou2023assistance}, which enables us to develop a two-step joint detector for both the PU and the SU to achieve a better decoding performance. For ease of exposition, we drop the time notation $l$ from now on.

\subsubsection{Detection at the PU}
Specifically, the joint detection at the PU is achieved through the use of the composite signal that consists of both the primary and secondary signals, defined as
\begin{align}
x_{\mathrm{pu}}(s,c)\triangleq \bm{h}_{p}^{H}\bm{w}s+\bm{g}_{p}^{H}\bm{\Theta}_{1}\bm{H}\bm{w}s+\bm{g}_{p}^{H}\bm{\Theta}_{2}\bm{H}\bm{w}sc,
\end{align}
where $x_{\mathrm{pu}}(s,c)\in \mathcal{A}_{x,p}$. Here, $\mathcal{A}_{x,p}$ denotes the constellation set of the primary composite signal $x_{\mathrm{pu}}(s,c)$, which can be obtained by using the channels, the constellation sets $\mathcal{A}_{s}$, $\mathcal{A}_{c}$, as well as $\bm{\Theta}_{1}$ and $\bm{\Theta}_{2}$.

With the primary composite signal, the received signal at the PU can be rewritten as
\begin{align}\label{eq: composite-signal-PU}
    y_{p}=x_{\mathrm{pu}}(s,c)+z_{p}.
\end{align}

From \eqref{eq: composite-signal-PU}, we can first decode the composite signal from the received signal at the PU using the classical maximum likelihood detection, and then map the primary composite signal into the primary and secondary signals. The whole procedures are summarized as follows.
\begin{align}
   \emph{ \text{Step 1:}}& \quad \hat{x}_{\mathrm{pu}}(\hat{s},\hat{c})\!=\!
    \arg \min_{x_{\mathrm{pu}}(s,c)\in\mathcal{A}_{x,p}}
    |y_{p}-x_{\mathrm{pu}}(s,c)|^2, \nonumber \\
    \emph{\text{Step 2:}}& \quad \quad \{\hat{s}, \hat{c}\} \xleftarrow{\text{mapping}} \hat{x}_{\mathrm{pu}}(\hat{s},\hat{c}). \label{eq:detec}
\end{align}

With the above two-step detection procedure, when the PU decodes $\hat{s}$ and $\hat{c}$, it takes $\hat{s}$ as the decoded primary signal. 
\subsubsection{Receiver Design at the SU}
Since the secondary signal $c$ is modulated over the primary signal $s$, decoding $c$ needs the information of $s$ for coherent detection. Thus, to pursue a better performance, we also adopt joint detection at the SU. Similarly, the above two-step joint detection method can be directly applied here. To be specific, the composite signal received at the SU is defined as
\begin{align}
    x_{\mathrm{su}}(s,c) \triangleq \bm{h}_{s}^{H}\bm{w}s+\bm{g}_{s}^{H}\bm{\Theta}_{1}\bm{H}\bm{w}s+\bm{g}_{s}^{H}\bm{\Theta}_{2}\bm{H}\bm{w}sc.
\end{align}

With the two-step detector proposed in \eqref{eq:detec}, when the SU decodes the primary and secondary signals $\hat{s}$ and $\hat{c}$, it takes $\hat{c}$ as its decoded secondary signal.
\section{BER Performance Analysis} \label{sec: performance-analysis}
In this section, we first build up the framework of BER performance analysis, which is applicable to both our proposed modulation scheme and its conventional counterpart.
Then, based on the performance analysis, we reveal the detection ambiguity and error-floor problems when using the conventional modulation scheme and we also show how to address the two problems by using our proposed scheme.
\subsection{BER Performance of the Primary System}
Under the proposed detector, the union bound~\cite{proakis2001digital} on the BER for decoding $s$ at the PU can be written as
\begin{align}
        P_{s}=&
       \sum_{i=1}^{|\mathcal{A}_{s}|} \sum_{m=1}^{|\mathcal{A}_{c}|}\sum_{j=1,j\neq i}^{|\mathcal{A}_{s}|} \sum_{k=1}^{|\mathcal{A}_{c}|} P(x_{\mathrm{pu}}(s_{i},c_{m})\rightarrow x_{\mathrm{pu}}(s_{j},c_{k})) 
        \nonumber \\  & \quad \quad \quad \quad \quad \quad \quad \quad \quad
       \times  \frac{1}{|\mathcal{A}_{s}||\mathcal{A}_{c}|}\frac{e(s_{i}\rightarrow s_{j})}{\log_{2}(|\mathcal{A}_{s}|)}, \label{eq: Ps}
\end{align}
where $e(s_{i}\rightarrow s_{j})$ denotes the Hamming distance between $s_{i}$ and $s_{j}$; $|\mathcal{A}_{s}|$ and $|\mathcal{A}_{c}|$ denote the cardinality of the sets $\mathcal{A}_{s}$ and $\mathcal{A}_{c}$; $P(x_{\mathrm{pu}}(s_{i},c_{m})\rightarrow x_{\mathrm{pu}}(s_{j},c_{k}))$ denotes the pairwise error probability between $x_{\mathrm{pu}}(s_{j},c_{k})$ and $x_{\mathrm{pu}}(s_{i},c_{m})$, given by
\begin{align}
    &P(x_{\mathrm{pu}}(s_{i},c_{m})\rightarrow x_{\mathrm{pu}}(s_{j},c_{k}))\nonumber\\
    &\!=\!P(|y_{p}-x_{\mathrm{pu}}(s_{i},c_{m})|^2\!>\!|y_{p}-{x}_{\mathrm{pu}}(s_{j},c_{k})|^2)
   \nonumber \\&\!=\!\mathcal{Q}\left(\sqrt{\frac{d_{\mathrm{pu}}((s_{i},c_{m}),(s_{j},c_{k}))}{2\sigma_{p}^{2}}}\right),
\end{align}
where 
\begin{align}
    d_{\mathrm{pu}}((s_{i},c_{m}),(s_{j},c_{k}))\!\triangleq\!|x_{\mathrm{pu}}(s_{i},c_{m})- x_{\mathrm{pu}}(s_{j},c_{k})|^2, \label{eq: distance-primary}
\end{align}
denotes the squared Euclidean distance between the two constellations of the primary composite signal, i.e.,  $x_{\mathrm{pu}}(s_{i},c_{m})$ and $x_{\mathrm{pu}}(s_{j},c_{k})$; $\mathcal{Q}(\cdot)$ denotes the Gaussian $\mathcal{Q}$-function.
\subsection{BER Performance of the Secondary System}
Similarly, the union bound on the BER for decoding $c$ at the SU can be written as
\begin{align}
    P_{c}=&
       \sum_{i=1}^{|\mathcal{A}_{s}|} \sum_{m=1}^{|\mathcal{A}_{c}|}\sum_{j=1}^{|\mathcal{A}_{s}|} \sum_{k=1,k\neq m}^{|\mathcal{A}_{c}|} \mathcal{Q}\left(\sqrt{\frac{d_{\mathrm{su}}((s_{i},c_{m}),(s_{j},c_{k}))}{2\sigma_{s}^{2}}}\right)\nonumber\\  &\quad \quad \quad \quad \quad \quad \quad \quad \quad \times  \frac{1}{|\mathcal{A}_{s}||\mathcal{A}_{c}|}\frac{e(c_{m}\rightarrow c_{k})}{\log_{2}(|\mathcal{A}_{c}|)}, \label{eq: Pc}
\end{align}
where 
\begin{align}
d_{\mathrm{su}}((s_{i},c_{m}),(s_{j},c_{k}))\!\triangleq\!|x_{\mathrm{su}}(s_{i},c_{m})\!-\! x_{\mathrm{su}}(s_{j},c_{k})|^2,
\end{align} 
denotes the squared Euclidean distance between the two constellations of the secondary composite signal, namely,  $x_{\mathrm{su}}(s_{i},c_{m})$ and $x_{\mathrm{su}}(s_{j},c_{k})$;  $e(c_{m}\!\rightarrow \! c_{k})$ denotes the Hamming distance between $c_{m}$ and $c_{k}$.
\subsection{BER Approximation via Minimum Euclidean Distance} \label{sec-distance}
Although the BER expressions are analyzed in \eqref{eq: Ps} and \eqref{eq: Pc}, they are very complicated to be optimized, which motivates us to find a simple but effective approximation of $P_{s}$ and $P_{c}$. Fortunately, it is observed that $P_s$ and $P_c$ are related to the Euclidean distances of the composite signals received at the PU and SU, respectively. Since the $\mathcal{Q}$-function is a monotonically decreasing function, a larger Euclidean distance indicates a lower BER~\cite{proakis2001digital}. Thus, we can obtain the upper bound of $P_{s}$ and $P_{c}$ through the use of minimum Euclidean distance, where the details can be found in Appendix \ref{appendix-A}.

Here, the minimum Euclidean distances of the composite signals received at the PU and the SU are defined as
\begin{align} \label{eq:minimum-Euclidean-PU}
    D_{\mathrm{pu}}\triangleq\min_{\substack{1\leq i,j \leq |\mathcal{A}_{s}|,i\neq j\\ 1\leq m,k \leq |\mathcal{A}_{c}|}} \{d_{\mathrm{pu}}((s_{i},c_{m}),(s_{j},c_{k}))\},
\end{align}
and
\begin{align} \label{eq:minimum-Euclidean-SU}
 D_{\mathrm{su}}\triangleq\min_{\substack{ 1\leq m,k \leq |\mathcal{A}_{c}|, m \neq k\\1\leq i,j \leq |\mathcal{A}_{s}|}}\{d_{\mathrm{su}}((s_{i},c_{m}),(s_{j},c_{k}))\},
\end{align}
respectively.

Note that for the PU, we focus on the BER performance of decoding $s$. Thus, when calculating the minimum Euclidean distance of the two constellations of the primary composite signal, i.e., $x_{\mathrm{pu}}(s_{i},c_{m})$ and $x_{\mathrm{pu}}(s_{j},c_{k})$ at the PU, it is required that $s_{i} \neq s_{j}$ regardless of the relationship between $c_{m}$ and $c_{k}$. This is because for the case $s_{i}=s_{j}$, even though we make a decision of $x_{\mathrm{pu}}(s_{j},c_{k})$ when the transmitted symbol is $x_{\mathrm{pu}}(s_{i},c_{m})$, the decoding of the primary signal $s$ at the PU is still correct. Besides, from a mathematical perspective, for the case $s_{i}=s_{j}$, we have $e(s_{i}\rightarrow s_{j})=0$. In this case, according to \eqref{eq: Ps}, the corresponding  $\mathcal{Q}$-function does not contribute to the BER calculation.
Likewise, for the secondary system, we focus on the BER performance of decoding $c$, and thus we require that $c_{m}\neq c_{k}$, i.e., $m\neq k$.
\subsection{Performance Limits of Existing Modulation Scheme} \label{sec:performance}
Based on the above analysis framework, we can also characterize the BER performance under the existing modulation scheme, where the mapping rule between the secondary signal and the phase-shift matrix is given by $\bm{\Theta}(c)=c\bm{\Theta}_{2}$. It can be viewed as a special case of our proposed scheme where $\bm{\Theta}_{1}=\bm{0}$.

Similarly, the BER performance of the PU and the SU under the existing scheme can be characterized by the minimum Euclidean distance. Taking the primary system as an example, the Euclidean distances of the primary composite signal can be classified into three types
\begin{align} \label{eq: distance-primary-conventional}
	\begin{cases}
	d_{1}\!=\!|\bm{\theta}_{2}^{H}\bm{F}_{p}\bm{w}|^2|c_{m}-c_{k}|^2, \quad  \  s_{i}=s_{j},c_{m}\neq c_{k}\\
	d_{2}\!=\!|\bm{h}_{p}^{H}\bm{w}|^2|s_{i}-s_{j}|^2, s_{i}c_{m}\!=\!s_{j}c_{k},s_{i}\!\neq\! s_{j}, c_{m}\!\neq\! c_{k} \\
	d_{3}\!=\!\left|\bm{h}_{p}^{H}\bm{w}\!+\!\bm{\theta}_{2}^{H}\bm{F}_{p}\bm{w}\frac{s_{i}-s_{j}\frac{c_k}{c_m}}{s_{i}\!-\!s_{j}}c_m\right|^2|s_{i}\!-\!s_{j}|^2,  \mathrm{otherwise}.
\end{cases} 
\end{align}
where $\bm{F}_{p}\triangleq\mathrm{diag}(\bm{g}_{p}^{H})\bm{H}$, $\bm{\theta}_{2}^{H}$ is the row vector form of the diagonal elements of $\bm{\Theta}_{2}$.

From \eqref{eq: distance-primary-conventional}, it is observed that in the second case where $s_{i}c_{m}\!=\!s_{j}c_{k}$, the Euclidean distance is independent of $\bm{\theta}_{2}$. Through this observation, we reveal two fundamental problems under the existing modulation scheme, which are shown in the following.

\textbf{\emph{Fundamental problem 1 (Detection ambiguity)}}: When the direct link is severely blocked by obstacles, i.e.,  $\bm{h}_{p}=\bm{0}$, the ambiguity problem arises~\cite{zhou2023modulation}. In this case, we have $d_{2}=0$ in \eqref{eq: distance-primary-conventional}. Physically speaking, the ambiguity means that there exist two sets of optimal estimators $\{s_{i},c_{m}\}$ and $\{s_{j},c_{k}\}$, which cannot be distinguished when they satisfy the condition $s_{i}c_{m}=s_{j}c_{k}$. Thus, the BER of the primary system $P_{s}=\mathcal{Q}(\sqrt{\frac{d_{2}}{2\sigma_{p}^{2}}})=0.5$ since $d_2=0$ when the direct link is blocked.

\textbf{\emph{Fundamental problem 2 (BER error-floor)}}: When the direct link is much weaker than the reflecting link, the error-floor problem arises. As we can see, given the channels and transmit beamforming vector $\bm{w}$, the second distance term $d_{2}=|\bm{h}_{p}^{H}\bm{w}|^2|s_{i}-s_{j}|^2$ in \eqref{eq: distance-primary-conventional} will become the minimum when the number of RIS reflecting elements is large. In this case, $d_{2}$ is a fixed value regardless of how large $N$ is. With the increase of $N$, the BER will be limited by this distance term $d_{2}$ and gradually approaches to the error floor, i.e., $\mathcal{Q}(\sqrt{\frac{d_{2}}{2\sigma_{p}^{2}}})$.
Note that the two problems hold for the secondary system as well. Here, we take the primary system as an example to demonstrate them.

Against this background, our proposed scheme could address the two fundamental problems due to the introduction of the assistance beamforming matrix $\bm{\Theta}_{1}$. On one hand, when the direct link is blocked, the reflecting link via $\bm{\Theta}_{1}$ can provide a virtual direct link to address the ambiguity problem. On the other hand, when the direct link is weaker than the reflecting link via RIS, $\bm{\Theta}_{1}$ can also help enhance the weak direct link to address the error-floor problem. From a mathematical perspective, when $\bm{\Theta}_{1}$ is introduced, the second distance term in \eqref{eq: distance-primary-conventional} becomes $d_{2}=|\bm{h}_{p}^{H}\bm{w}+\bm{\theta}_{1}\bm{F}_{p}\bm{w}|^2|s_{i}-s_{j}|^2$ that can be enlarged with the increase of number of reflecting elements, thereby addressing the two problems.
With the proposed modulation scheme, the BER of the primary and secondary systems are dependent on the assistance and transmission beamforming matrices, which should be jointly designed to satisfy different performance requirements of the primary and secondary systems, shown in the next section.

% Intuitively, it is observed that when the direct link of the primary system is blocked, the introduction of $\bm{\Theta}_{1}$ can serve as a virtual direct link to support the primary system. At the same time, $\bm{\Theta}_{1}$ can also help the secondary system to eliminate the ambiguity problem in the presence of direct link

% It can be seen that, when the direct link is blocked, the ambiguity problem arises. When we increases, the their exists the error floor. And, the similar results can be made for the case at the SU.
% Before proceeding the phase shifts optimization of our proposed RIS design scheme, we first revisit the performance limitations of the conventional RIS design scheme in SR. Note that the conventional RIS design can be viewed as a special case of our proposed RIS design with $\bm{\Phi}=c\bm{\Phi}_{2}$ and $\bm{\Phi}_{1}=\bm{0}$. Therefore, the BER performance under the conventional RIS design can be similarly characterized by the minimum Euclidean distance defined in \eqref{eq:minimum-Euclidean-PU} and \eqref{eq:minimum-Euclidean-SU}. 

% It is not difficult to see that when $\bm{\Phi}(c)=c\bm{\Phi}_{2}$, there exists one distance term becomes $2|h_{p}|$ which is independent of $\bm{\Phi_{2}}$. As we can see, if we increase the number of RIS reflecting elements, the distance term $2|h_{p}|$ and $2|h_{s}|$ will become the minimum Euclidean distance terms at the PU and the SU, respectively. That is to say, the BER performance of PU and the SU will be limited by this distance regardless of $N$.
\section{Problem Formulation} \label{sec:problem-formulation}
The above BER approximation in Sec. \ref{sec-distance} provides us with an alternative way to optimize the BER performance via the defined minimum Euclidean distances $D_{\mathrm{pu}}$ and $D_\mathrm{su}$. In this section, we aim to maximize the minimum Euclidean distance of the secondary composite signal $D_{\mathrm{su}}$ subject to the requirement of the minimum Euclidean distance of the primary composite signal $D_{\mathrm{pu}}$. Here, to guarantee the performance enhancement of the primary system as compared to the case without RIS, the performance requirement of the primary system can be adjusted based on the following two special schemes.  
% The above BER approximation provides us an alternative way to optimize the BER performance via the defined minimum Euclidean distances $D_p$ and $D_s$. Thus, in this section, we aim to maximize the minimum Euclidean distance of the secondary composite signal $D_{s}$ subject to the requirement of the minimum Euclidean distance of the primary composite signal $D_{p}$ and the modulus constraints of the RIS phase-shift matrix $\bm{\Theta}$. Specifically, the requirement of $D_{p}$ can be adjusted based on the following two special schemes. 
\begin{itemize}
    \item \emph{Scheme I (RIS is off)}:  In this case, we have $\bm{\Theta}\!=\!\bm{0}$ and there exists the primary system only. Obviously, the optimal transmit beamforming vector in this case is maximum-ratio-transmission (MRT), given by 
    $\bm{w}^{I}=\sqrt{P_t}\frac{\bm{h}_{p}}{\Vert\bm{h}_p\Vert}$, where $P_{t}$ denotes the maximum transmit power at the PT.
    In this case, the 
    minimum Euclidean distance is given by
    \begin{align}
        D^{I}&\triangleq \min_{\substack{1\leq i,j \leq |\mathcal{A}_{s}|,i\neq j}} |\bm{h}_{p}^{H}\bm{w}^{I}(s_{i}-s_{j})|^2. \label{eq: special-scheme-1}
    \end{align}
    \item \emph{Scheme II (RIS purely assists the primary system)}: In this case, we have $\bm{\Theta}\!=\!\bm{\Theta}_{1}$ with $\bm{\Theta}_{2}=\bm{0}$. This degrades to the classical RIS-aided MISO system~\cite{wu2019intelligent}, where the optimal transmit beamforming vector $\bm{w}^{II}$ and the phase shifts $\bm{\Theta}^{II}$ can be iteratively obtained with closed-form solutions (Iterative details can be found in~\cite{wu2019intelligent}). In this case, the minimum Euclidean distance is given by
    \begin{align} \label{eq: special-scheme-2}
    D^{II}\!\triangleq \!  \min_{\substack{1\leq i,j \leq |\mathcal{A}_{s}|\\i\neq j}} \left|\left(\bm{h}_{p}^{H}\bm{w}^{II}\!+\!\bm{g}_{p}^{H}\bm{\Theta}^{II}\bm{H}\bm{w}^{II}\right)(s_{i}\!-\!s_{j})\right|^2.
    \end{align}
\end{itemize}

Based on the above two special schemes, the minimum Euclidean distance $D_{\mathrm{pu}}$ can be varied from $D^{I}$ to $D^{II}$ by jointly designing $\bm{w}$, $\bm{\Theta}_{1}$ and $\bm{\Theta}_{2}$. Here, we introduce an assistance factor $\delta \in [0,1]$ to control the performance requirement of the primary transmission, and then formulate the mathematical problem as follows
\begin{subequations}
	\begin{align}
	\underline{\textbf{\text{P1:}}} \  
	&\max \limits_{\bm{w},\bm{\Theta}_{1},\bm{\Theta}_{2}}\ D_{\mathrm{su}}\\
	&\ \mbox{s.t.} \  \left|[\bm{\Theta}_{1}\!+\!\bm{\Theta}_{2}c]_{n,n}\right|\leq 1, \forall n\!=\!1,\cdots,N, c\in \mathcal{A}_{c}, \label{eq:P1-b} \\
	& \quad \quad  D_{\mathrm{pu}}\geq \eta, \label{eq:P1-c}\\
    & \quad \quad \Vert \bm{w} \Vert^2\leq P_t, \label{eq:P1-d}
    \\
	& \quad \quad \eqref{eq:minimum-Euclidean-PU}, \eqref{eq:minimum-Euclidean-SU}. \nonumber
	\end{align}
\end{subequations}
where $\eta = D^{I}+\delta (D^{II}-D^{I})$ denotes the distance requirement controlled by $\delta$; \eqref{eq:P1-b} denotes the modulus constraint of the phase shift associated with each reflecting element and each $c\! \in \! \mathcal{A}_{c}$. Under this constraint, the amplitude of each diagonal element of $\bm{\Theta}_{1}$ and $\bm{\Theta}_{2}$ may not be unity; \eqref{eq:P1-d} is the maximum transmit power constraint.

In \eqref{eq:P1-c}, $D^{II}-D^{I}$ denotes the maximum performance gain that the primary system can achieve compared to the case without RIS. If $\delta\!=\!1$, this requires the RIS need to purely assist the primary system without transmitting the secondary signal. If $\delta\!=\!0$, this implies that when RIS transmits the secondary signal by sharing the spectrum of the primary one, it needs to guarantee that the primary system can benefit from such sharing or at least does not degrade its performance. 

Generally speaking, it is difficult to solve \textbf{P1} due to the following reasons. First, the modulus phase shift constraints involve the optimization of two matrices. This form is different from the existing phase shift constraint that only requires optimizing one phase shift matrix~\cite{zhou2022cooperative,wang2021intelligent,hua2021novel}. Thus, the conventional phase shifts optimization method cannot be directly applied. Second, due to the non-convex objective function and constraints, there is no standard method to obtain the optimal solution directly.
% \vspace{-0.1cm}
\section{Proposed Algorithms}
\label{sec: proposed-algorithm}
% \vspace{-0.1cm}
In this section, we aim to solve the problem \textbf{P1}. Although the formulated problem is challenging to solve, the widely used alternating optimization (AO) algorithm offers us an efficient way to solve it. The basic idea is to decouple the original problem into two sub-problems with respect to (w.r.t.) $\bm{w}$ and $\{\bm{\Theta}_{1},\bm{\Theta}_{2}\}$, which are solved iteratively until the convergence is met. 

By applying the AO algorithm, the original problem \textbf{P1} can be divided into the following two sub-problems
\begin{subequations}
	\begin{align}
	\underline{\textbf{\text{P1-A:}}} \  
	&\max \limits_{\bm{w}}\ D_{\mathrm{su}}\\
	&\ \mbox{s.t.} 
\ \eqref{eq:minimum-Euclidean-PU}, \eqref{eq:minimum-Euclidean-SU}, \eqref{eq:P1-c}, \eqref{eq:P1-d}.\nonumber
	\end{align}
\end{subequations}
and 
\begin{subequations}
	\begin{align}
	\underline{\textbf{\text{P1-B:}}} \  
	&\max \limits_{\bm{\Theta}_{1},\bm{\Theta}_{2}}\ D_{\mathrm{su}}\\
	&\ \mbox{s.t.} 
\ \eqref{eq:minimum-Euclidean-PU}, \eqref{eq:minimum-Euclidean-SU}, \eqref{eq:P1-b}, \eqref{eq:P1-c}.\nonumber
	\end{align}
\end{subequations}

Here, \textbf{P1-A} is related to the active transmit beamforming optimization, while \textbf{P1-B} is related to assistance and transmission beamforming optimization. Next, we will solve the two sub-problems, respectively.
\subsection{Active Transmit Beamforming Optimization}
Given the assistance and transmission beamforming matrices $\bm{\Theta}_{1}$ and $\bm{\Theta}_{2}$, we aim to optimize the active transmit beamforming vector $\bm{w}$. To do so, we first transform the minimum Euclidean distance into a tractable form. Specifically, the Euclidean distance of the primary composite signal can be written as a function of $\bm{w}$, given by
\begin{align} 
    d_{\mathrm{pu}}((s_{i},c_{m}), (s_{j},c_{k})) =|\bm{a}_{p}^{H}\bm{w}|^2,
     \label{eq: quadratic-form-primary}
\end{align}
where 
\begin{align} \nonumber
\bm{a}_{p}&\triangleq((\bm{h}_{p}^{H}\!+\!\bm{g}_{p}^{H}\bm{\Theta}_{1}\bm{H})(s_{i}-s_{j})\! +\!\bm{g}_{p}^{H}\bm{\Theta}_{2}\bm{H}(s_{i}c_{m}\!-\!s_{j}c_{k}))^{H}.
\end{align}

Similarly, the Euclidean distance of the secondary composite signal can be rewritten as $d_{\mathrm{su}}((s_{i},c_{m}), (s_{j},c_{k}))=|\bm{a}_{s}^{H}\bm{w}|^2$, where $\bm{a}_{s}\triangleq((\bm{h}_{s}^{H}+\bm{g}_{s}^{H}\bm{\Theta}_{1}\bm{H})(s_{i}-s_{j})+\bm{g}_{s}^{H}\bm{\Theta}_{2}\bm{H}(s_{i}c_{m}-s_{j}c_{k}))^{H}$.

Then, by denoting $\bm{W}=\bm{w}\bm{w}^{H}$ and introducing an auxiliary variable $t_A$ that is non-negative, the sub-problem \textbf{P1-A} can be transformed into a rank-constrained semi-definite programming (SDP) problem, given by
\begin{subequations} 
\begin{align}
\underline{\textbf{\text{P1-A1:}}} \  
&\max \limits_{\bm{W},t_A}\ t_A\\
&\ \mbox{s.t.} \ \mathrm{Tr}(\bm{W})\leq P_t, \label{eq:P1-A1-b}
\\ &\quad \quad \mathrm{Tr}(\bm{W}\bm{A}_{p})\geq \eta,\forall s_{i} \neq s_{j},    \\
& \quad \quad \mathrm{Tr}(\bm{W}\bm{A}_{s})\geq t_A, \forall c_{m} \neq c_{k},\\
& \quad \quad \bm{W}\succeq0,\ t_A\geq0,\\
& \quad \quad \mathrm{rank}(\bm{W})=1,
\end{align}
\end{subequations}
where $\bm{A}_{p}=\bm{a}_{p}\bm{a}_{p}^{H}$ and $\bm{A}_{s}=\bm{a}_{s}\bm{a}_{s}^{H}$. In \textbf{P1-A1}, it is observed that the power budget constraint \eqref{eq:P1-A1-b} is active for the optimal solution $\bm{W}^{\star}$. This can be proved by contradiction. Assume the optimal beamforming vector is denoted as $\bm{w}^{\star}=\sqrt{P}_{0}\bm{w}_{0}$ where $\Vert \bm{w}_{0}\Vert=1$ and $P_{0}<P_{t}$. In this case, we can always increase the transmit power to $P_{0}^{'}$ that satisfies $ P_{0}<P_{0}^{'}\leq P_{t}$, and obtain another beamforming vector, $\sqrt{P_{0}^{'}}\bm{w}_{0}$, which leads to a larger value of objective function. However, this contradicts the assumption where $\sqrt{P_{0}}\bm{w}_{0}$ is optimal solution. Thus, the optimal $\bm{W}^{\star}$ is obtained with $\mathrm{Tr}(\bm{W}^{\star})=P_{t}$.

For \textbf{P1-A1}, both the objective function and constraints are convex except for the rank-one constraint. One common method is to use the semi-definite relaxation (SDR) technique to drop the rank-one constraint, followed by the Gaussian randomization method. However, it can be similarly proved that optimal $\bm{W}^{\star}$ to \textbf{P1-A1} is of rank-one, according to the results of \cite{zuo2022joint}. Thus,  the relaxation of the rank-one constraint is tight and there is no need to perform the Gaussian randomization. After obtaining $\bm{W}^{\star}$, we can recover optimal $\bm{w}^{\star}$ by directly performing singular value decomposition of $\bm{W}^{\star}$, given by $\bm{U}\bm{\Sigma}\bm{{U}}^{H}$, where $\bm{U} \in \mathbb{C}^{M \times M}$ and $\bm{\Sigma}\in \mathbb{C}^{M \times M}$ are a unitary matrix and a diagonal matrix composed of the singular value of $\bm{W}^{\star}$, respectively. Finally, $\bm{w}^{\star}$ is obtained as $\bm{w}^{\star} = \sqrt{[\bm{\Sigma}]_{1,1}}\bm{U}[:,1]$, where $[\bm{\Sigma}]_{1,1}$ denotes the first diagonal element of $\bm{\Sigma}$ and $\bm{U}[:,1]$ denotes the first column of $\bm{U}$.

% using which the problem becomes a convex one. Then, it can be efficiently solved using the existing optimization tools, e.g., CVX.  

% Note that it can be readily proved that the obtained transmit beamforming matrix is always rank-one. Thus, Gaussian randomization is not needed and we can directly perform singular value decomposition to obtain the rank-one solution as $\bm{w}^{(l)}$.
\subsection{Assistance and Transmission Beamforming Optimization}
Given the transmit beamforming vector $\bm{w}$, we next aim to optimize the assistance and transmission beamforming matrices, whose vector forms are denoted by  $\bm{\theta}_{1}\!=\![\theta_{1,1},\cdots,\theta_{1,N}]^{H}\in \mathbb{C}^{N\times 1}$, $\bm{\theta}_{2}\!=\![\theta_{2,1},\cdots,\theta_{2,N}]^{H}\in \mathbb{C}^{N\times 1}$. In the following, we first combine the two optimization variables $\bm{\theta}_{1}$ and $\bm{\theta}_{2}$ as a single variable, and then transform it into a quadratically constrained quadratic program (QCQP) problem that can be solved by the well-designed SCA technique.

The phase-shift vector of RIS can be expressed as
\begin{align}
\bm{\theta}
\triangleq\bm{b}(c) \bm{\Phi},
% =
% \begin{pmatrix}
% \theta_{1,1} & \theta_{1,2} & \cdots & \theta_{1,N}\\
% \theta_{2,1} & \theta_{2,2} & \cdots & \theta_{2,N}
% \end{pmatrix}
\end{align}
where $\bm{b}(c)\!=\!\begin{pmatrix}
1 & c
\end{pmatrix}\in \mathbb{C}^{1\times2}$, $\bm{\Phi}\!=\!\begin{pmatrix}
\bm{\theta}_{1}^{H}\\
\bm{\theta}_{2}^{H}
\end{pmatrix}\in\mathbb{C}^{2\times N}$ is the variable that combines the assistance and transmission beamforming vectors.

Accordingly, the modulus constraints of the phase-shift vector $\bm{\theta}$, i.e., \eqref{eq:P1-b}, can be rewritten as
\begin{align} \label{eq: overall-phase-shifts}
    |\bm{b}(c)\bm{\Phi}\bm{e}_{n}| \leq1, \forall n=1,\cdots,N, c\in \mathcal{A}_{c},
\end{align}
where $\bm{e}_{n}\in\mathbb{C}^{N\times 1}$ is the canonical basis vector whose $n$-th element is $1$ while the other elements are $0$.
Moreover, the composite signals at the PU and SU can be rewritten as
\begin{align}
   x_{\mathrm{pu}}(s,c)&=\bm{h}_{p}^{H}\bm{w}s+s\bm{b}(c)\bm{\Phi}\bm{F}_{p}\bm{w}, \\
    x_{\mathrm{su}}(s,c)&=\bm{h}_{s}^{H}\bm{w}s+s\bm{b}(c)\bm{\Phi}\bm{F}_{s}\bm{w}.
\end{align}
where $\bm{F}_{p}\triangleq\mathrm{diag}(\bm{g}_{p}^{H})\bm{H}$, $\bm{F}_{s}\triangleq \mathrm{diag}(\bm{g}_{s}^{H})\bm{H}$.

With the above transformations, we can recast the sub-problem \textbf{P1-B} into the following \textbf{P1-B1}, where the optimization variable becomes $\bm{\Phi}$
\begin{subequations}
\begin{align}
\underline{\textbf{\text{P1-B1:}}} \  
&\max \limits_{\bm{\Phi}}\ D_{\mathrm{su}}\\
&\ \mbox{s.t.} \  
D_{\mathrm{pu}}\!=\!\min\{|(s_{i}\bm{b}(c_{m})\!-\!s_{j}\bm{b}(c_{k}))\bm{\Phi}\bm{F}_{p}\bm{w} \nonumber\\
& \quad  \quad \quad \quad +\bm{h}_{p}^{H}\bm{w}(s_{i}-s_{j})|^2\},\forall s_{i} \neq s_{j},\\
&\quad \quad D_{\mathrm{su}}\!=\!\min\{ |(s_{i}\bm{b}(c_{m})\!-\!s_{j}\bm{b}(c_{k}))\bm{\Phi}\bm{F}_{s}\bm{w} \nonumber\\
& \quad  \quad \quad \quad + \bm{h}_{s}^{H}\bm{w}(s_{i}-s_{j})|^2\}, \forall c_{m}\neq c_{k},\\
& \quad \quad \eqref{eq:P1-c}, \eqref{eq: overall-phase-shifts}, \nonumber
\end{align}
\end{subequations}

Next, we rewrite the Euclidean distance in \textbf{P1-B1} to recast it as a quadratic function. Let $\bm{a}\triangleq s_{i}\bm{b}(c_{m})\!-\!s_{j}\bm{b}(c_{k})$, then the Euclidean distance between the primary composite signal $x_{\mathrm{pu}}(s_{i},c_{m})$ and $x_{\mathrm{pu}}(s_{j},c_{k})$ is given by 
\begin{align} 
    &d_{\mathrm{pu}}((s_{i},c_{m}), (s_{j},c_{k})) \nonumber\\
    &= |\bm{h}_{p}^{H}\bm{w}(s_{i}-s_{j})|^2+|\bm{a}\bm{\Phi}\bm{F}_{p}\bm{w}|^2 \nonumber \\
   & \quad \quad  \quad \quad \quad \quad \quad  + 2\Re\{(\bm{a}\bm{\Phi}\bm{F}_{p}\bm{w})^{H}\bm{h}_{p}^{H}\bm{w}(s_{i}-s_{j})\} . \label{eq: quadratic-primary-distance-1}
\end{align}

In \eqref{eq: quadratic-primary-distance-1}, the term $|\bm{a}\bm{\Phi}\bm{F}_{p}\bm{w}|^2$ can be further derived as
\begin{align}
    |\bm{a}\bm{\Phi}\bm{F}_{p}\bm{w}|^2 & = \mathrm{Tr}(\bm{\Phi}^{H}\bm{a}^{H}\bm{a}\bm{\Phi}\bm{F}_{p}\bm{W}\bm{F}_{p}^{H}) \nonumber\\
   &\overset{(a)}{=} \mathrm{vec}(\bm{\Phi})^{H}\mathrm{vec}(\bm{a}^{H}\bm{a}\bm{\Phi}\bm{F}_{p}\bm{W}\bm{F}_{p}^{H})\nonumber\\
   & \overset{(b)}{=}\! \bm{v}^{H}\bm{L}_{p} \bm{v} \label{eq: quadratic-primary-distance-2}
\end{align}
where $\bm{W}=\bm{w}\bm{w}^{H}$;
$(a)$ holds due to the fact $\mathrm{Tr}(\bm{X}^{H}\bm{Y})\!=\!\mathrm{vec}(\bm{X})^{H}\mathrm{vec}(\bm{Y})$;  $(b)$ follows from $\mathrm{vec}(\bm{X}\bm{Y}\bm{Z})\!=\!(\bm{Z}^{T}\otimes \bm{X})\mathrm{vec}(\bm{Y})$ and the variable definitions of $\bm{v}\! \triangleq \! \mathrm{vec}(\bm{\Phi})\in\mathbb{C}^{2N\times 1}$, $\bm{L}_{p}\!\triangleq \! (\bm{F}_{p}\bm{W}\bm{F}_{p}^{H})^{T}\otimes \bm{a}^{H}\bm{a}$; 

With the defined $\bm{v}$, the third term in \eqref{eq: quadratic-primary-distance-1} and the modulus constraint in \eqref{eq: overall-phase-shifts} can be derived as
\begin{align}
    &2\Re\{(\bm{a}\bm{\Phi}\bm{F}_{p}\bm{w})^{H}\bm{h}_{p}^{H}\bm{w}(s_{i}-s_{j})\}\triangleq 2\Re\{\bm{v}^{H}\bm{l}_{p}\}, \label{eq: quadratic-primary-distance-3}\\
    &|\bm{b}(c)\bm{\Phi}\bm{e}_{n}|^2 \triangleq \bm{v}^{H} (\bm{e}_{n}\bm{e}_{n}^{H}\otimes \bm{b}(c)^{H}\bm{b}(c))\bm{v}, \label{eq: overall-phase-shifts-new}
\end{align}
where $\bm{l}_{p}=\mathrm{vec}(\bm{a}^{H}\bm{h}_{p}^{H}\bm{W}\bm{F}_{p}^{H}(s_{i}-s_{j}))$. 

Similar to \eqref{eq: quadratic-primary-distance-1}, \eqref{eq: quadratic-primary-distance-2}, and \eqref{eq: quadratic-primary-distance-3}, we can also rewrite the Euclidean distance of the secondary composite signal as a quadratic function of $\bm{v}$ and reformulate \textbf{P1-B1} as \textbf{P1-B2}, where the optimization variable now becomes $\bm{v}=\mathrm{vec}(\bm{\Phi})$, given by
\begin{subequations}
	\begin{align}
	\underline{\textbf{\text{P1-B2:}}} \  
	&\max \limits_{\bm{v}}\ D_{\mathrm{su}}\\
	&\ \mbox{s.t.} \  
	D_{\mathrm{pu}}\!=\!\min\{\bm{v}^{H}\bm{L}_{p} \bm{v}+2\Re\{\bm{v}^{H}\bm{l}_{p}\} \nonumber\\
	&\quad \quad \quad \quad  \quad \ +|\bm{h}_{p}^{H}\bm{w}(s_{i}-s_{j})|^2\},\forall s_{i} \neq s_{j}, \label{eq: P4-b} \\
	&\quad \quad  D_{\mathrm{su}}\!=\!\min\{\bm{v}^{H}\bm{L}_{s} \bm{v}+2\Re\{\bm{v}^{H}\bm{l}_{s}\} \nonumber\\
	&\quad \quad \quad \quad  \quad \ +|\bm{h}_{s}^{H}\bm{w}(s_{i}-s_{j})|^2\},\forall c_{m} \neq c_{k}, \label{eq: P4-c}\\
	&\quad \quad \bm{v}^{H} \bm{Q}_{n}\bm{v}\leq 1, \forall n,c \label{eq: P4-d}\\
    & \quad \quad \eqref{eq:P1-c}, \nonumber
% 	s_{i}, s_{j}\in\mathcal{A}_{s},c_{m}\neq c_{k},
	\end{align}
\end{subequations}
where we have
\begin{align}
    \bm{L}_{s}&\!=\! (\bm{F}_{s}\bm{W}\bm{F}_{s}^{H})^{T}\otimes \bm{a}^{H}\bm{a},\\
    \bm{l}_{s}&\!=\! \mathrm{vec}(\bm{a}^{H}\bm{h}_{s}^{H}\bm{W}\bm{F}_{s}^{H}(s_{i}-s_{j})),\\
    \bm{Q}_{n}&=\bm{e}_{n}\bm{e}_{n}^{H}\otimes \bm{b}(c)^{H}\bm{b}(c).
\end{align}

Through our transformations, \textbf{P1-B2} belongs to the class of QCQP problems. However, it is still difficult to solve \textbf{P1-B2} due to the non-convex constraints \eqref{eq:P1-c}, \eqref{eq: P4-d}, and objective function. To address the non-convexity, we resort to the classical SCA technique that is commonly used to solve the QCQP-related problems. Suppose we have a convex quadratic function of $f(\bm{v})=\bm{v}^{H}\bm{X}\bm{v}$ with $\bm{X}$ being a positive semi-definite matrix, then the SCA technique aims to find a surrogate function of $f(\bm{v})$ by taking the first-order Taylor expansion of $f(\bm{v})$ at any feasible point. Hence, given the point $\bm{v}_{q}$ at the $q$-th iteration, the convex function $f(\bm{v})$ can be globally lower-bounded by 
\begin{align}
\bm{v}^{H}\bm{X}\bm{v}\geq 2\Re\{\bm{v}_{q}^{H}\bm{X}\bm{v}\}-\bm{v}_{q}^{H}\bm{X}\bm{v}_{q},
\end{align}

Considering that both $\bm{L}_{p}$ and $\bm{L}_{s}$ are positive semi-definite matrices, SCA 
can be applied to transform the non-convex constraints \eqref{eq: P4-b} and \eqref{eq: P4-c} into linear ones. 
% Particularly, for the unit-modulus constraint \eqref{eq: P4-d}, it can be rewritten as $1\leq \bm{v}^{H} \bm{Q}_{n}\bm{v} \leq 1$. For the non-convex parts $\bm{v}^{H} \bm{Q}_{n}\bm{v} \geq 1$, it can be recast as $2\Re\{\bm{v}_{q}^{H}\bm{Q}_{n}\bm{v}\}-\bm{v}_{q}^{H}\bm{Q}_{n}\bm{v}_{q}\geq 1$ via SCA technique. 
By introducing an auxiliary variable $t_B$ that is non-negative, \textbf{P1-B2} can be transformed into \textbf{P1-B3}, given by
\begin{subequations}
	\begin{align}
	\underline{\textbf{\text{P1-B3:}}} \  
	&\max \limits_{\bm{v},t_B}\ t_B\\
	&\ \mbox{s.t.} \  
	2\Re\{(\bm{v}_{q}^{H}\bm{L}_{p}+\bm{l}_{p}^{H})\bm{v}\}+|\bm{h}_{p}^{H}\bm{w}(s_{i}-s_{j})|^2
	\nonumber\\
	& \quad \quad \quad \quad \quad-\bm{v}_{q}^{H}\bm{L}_{p}\bm{v}_{q} \geq \eta, \forall s_{i} \neq s_{j}, \label{eq:P5-b}
	\\
	&\quad \quad 2\Re\{(\bm{v}_{q}^{H}\bm{L}_{s}+\bm{l}_{s}^{H})\bm{v}\}+|\bm{h}_{s}^{H}\bm{w}(s_{i}-s_{j})|^2
	\nonumber\\
	&\quad \quad \quad \quad \quad -\bm{v}_{q}^{H}\bm{L}_{s}\bm{v}_{q} \geq t_B, \forall c_{m} \neq c_{k}, \label{eq:P5-c}
	\\
     & \quad \quad \bm{v}^{H} \bm{Q}_{n}\bm{v} \leq 1,\forall n\!=1, \cdots,N,       \\
	&\quad \quad t_B\geq0.
	\end{align}
\end{subequations}

In \textbf{P1-B3}, the constraint \eqref{eq:P5-b} combines the constraints \eqref{eq:P1-c} and \eqref{eq: P4-b} in \textbf{P1-B2}, which means that all the Euclidean distances of the primary composite signal should be larger than the distance requirement controlled by the assistance factor $\delta$. Similarly, the constraint \eqref{eq:P5-c} means that all the Euclidean distances of the secondary composite signal should be larger than $t_B$ and our objective is to maximize $t_B$. 
Then, \textbf{P1-B3} is a convex problem and thus can be efficiently solved by CVX~\cite{grant2014cvx}. For convenience, the overall algorithm for solving \textbf{P1} is summarized in Algorithm \ref{algorithm1}.
\begin{algorithm} [t] 
	\caption{Proposed Algorithm for Solving \textbf{P1}} 
	\label{algorithm1}  
	\begin{algorithmic}[1] 
		\STATE Initialize $\bm{w}^{(0)}$, $\bm{\Theta}_{1}^{(0)}=\mathrm{diag}(\bm{\theta}_{1}^{(0)})$, $\bm{\Theta}_{2}^{(0)}=\mathrm{diag}(\bm{\theta}_{2}^{(0)})$, $\bm{v}_{0}=\mathrm{vec}(\bm{\theta}_{1}^{(0)};\bm{\theta}_{2}^{(0)})$ and iteration number $q=0$.	\\
		\REPEAT 
        \STATE Given $\bm{\theta}_{1}^{(q)}$ and $\bm{\theta}_{2}^{(q)}$, obtain $\bm{w}^{(q+1)}$ by solving the sub-problem \textbf{P1-A1}, and set $D_{\mathrm{su}}^{(q)}=t_A$.\\
		\STATE Given $\bm{w}^{(q+1)}$ and $\bm{v}_{q}=\mathrm{vec}(\bm{\theta}_{1}^{(q)};\bm{\theta}_{2}^{(q)})$, obtain $\bm{v}_{q+1}$ by solving the sub-problem \textbf{P1-B3} and set $D_{\mathrm{su}}^{(q+1)}=t_B$. \\
		\STATE Update $\bm{\theta}_{1}^{(q+1)}$ and $\bm{\theta}_{2}^{(q+1)}$ from $\bm{v}_{q+1}$  and set $q =q+1$.\\
		\UNTIL{The objective function $D_{\mathrm{su}}^{(q)}$ converges.}
		\STATE Obtain the $\bm{w}^{\star}=\bm{w}^{(q)}$, $\bm{\Theta}_{1}^{\star}=\mathrm{diag}(\bm{\theta}_{1}^{(q)})$ and $\bm{\Theta}_{2}^{\star}=\mathrm{diag}(\bm{\theta}_{2}^{(q)})$.
	\end{algorithmic} 
\end{algorithm}
\section{Low-complexity Assistance and Transmission Beamforming Design}  
\label{sec:theorectical-analysis}
In the previous section, we have proposed algorithms to solve the formulated problem \textbf{P1}. To draw more insights and reduce the complexity, in this section, we propose a low-complexity assistance-transmission beamforming structure and analyze how to allocate the energy for assistance and transmission beamforming matrices under the modulus constraints.

For simplicity, we consider the scenario of a single-antenna PT. In this case, the channel coefficients of the PT-RIS link, the PT-PU link, and the PT-SU link become $\bm{h}$, $h_{p}$, $h_s$, respectively.
Recall that we have denoted the vector forms of the assistance and transmission beamforming matrices by $\bm{\theta}_{1}=[\theta_{1,1},\cdots,\theta_{1,N}]^{H}$, $\bm{\theta}_{2}=[\theta_{2,1},\cdots,\theta_{2,N}]^{H}$ in Sec. \ref{sec: proposed-algorithm}. Then, we have $\bm{g}_{p}^{H}\bm{\Theta}_{i}\bm{h}=\bm{\theta}_{i}^{H}\bm{f}_{p}$ and $\bm{g}_{s}^{H}\bm{\Theta}_{i}\bm{h}=\bm{\theta}_{i}^{H}\bm{f}_{s}$, $\forall i= 1,2$. Here, we let $\bm{f}_{p}=\mathrm{diag}(\bm{g}_{p}^{H})\bm{h}$ and $\bm{f}_{s}=\mathrm{diag}(\bm{g}_{s}^{H})\bm{h}$ be the primary and secondary reflecting channels, which 
refer to the PT-RIS-PU link and the PT-RIS-SU link, respectively. 
\subsection{Proposed Assistance-Transmission Beamforming Structure}
Besides the assistance beamforming vector $\bm{\theta}_{1}$ can directly provide assistance for the primary system, the transmission beamforming vector $\bm{\theta}_{2}$ can help enhance the primary system under the condition that the secondary signal can be decoded at the PU. Therefore, joint decoding is adopted at the PU to guarantee a better performance of the primary system. However, doing so will change the original receiver of the PU that only needs to decode the primary signal, and thus increases the decoding complexity at the PU. 

To preserve the assistance capability of RIS while eliminating the effect of secondary transmission on the primary system, we propose a low-complexity assistance-transmission beamforming structure by borrowing the idea from the classical MRT and ZF beamforming techniques. Specifically, the assistance beamforming vector is designed as the MRT of the primary reflecting channel, while the transmission beamforming vector is designed by projecting the secondary reflecting channel onto the null space of the primary reflecting channel.
By doing so, the primary system can still obtain assistance from the RIS while there is no need to decode the secondary signal at the PU.

However, due to the modulus constraints of RIS phase shifts, the above MRT and ZF operations need some modifications. The details are demonstrated in the following three steps. 
\subsubsection{Step 1 (MRT of $\bm{\theta}_{1}$ for Assistance)} We let the basis vector of $\bm{\theta}_{1}$ be ${\bm{\theta}}_{p}$. As ${\bm{\theta}}_{p}$ is used to assist the primary system, it can be obtained by solving the following problem.
\begin{subequations}
	\begin{align}
	\underline{\textbf{\text{P-MRT:}}} \  
	&\max \limits_{{\bm{\theta}}_{p}}\ |h_{p}+{\bm{\theta}}_{p}^{H}\bm{f}_{p}|^2\\
	&\ \mbox{s.t.} \ \  
|{\theta}_{p,n}|=1, \forall n=1,\cdots,N.
	\end{align}
\end{subequations}

Such a problem has been studied in~\cite{wu2019intelligent}, in which the optimal solution of ${\bm{\theta}}_{p}$ is given by 
\begin{align}
{\bm{\theta}}_{p}=\left[e^{\jmath(\angle (h_{p})-\angle(f_{p,1}))},\cdots,e^{\jmath(\angle (h_{p})-\angle(f_{p,N}))}\right]^{H}.
\end{align}

The above solution suggests that the phase shifts of $\bm{\theta}_{1}$ should be designed such that 
the signal going through the primary reflecting channel via $\bm{\theta}_{p}$ and the primary direct link channel $h_{p}$ can be coherently combined at the PU. 
\subsubsection{Step 2 (ZF of $\bm{\theta}_{2}$ for Transmission)}
Let the basis vector of $\bm{\theta}_{2}$ be ${\bm{\theta}}_{s}^{\perp}$.
As for ${\bm{\theta}}_{s}^{\perp}$, it should be designed to enhance the secondary reflecting channel while suppressing its impact on the primary reflecting channel, which can be obtained by solving the following problem.
\begin{subequations}
	\begin{align}
	\underline{\textbf{\text{P-ZF:}}} \  
	&\max \limits_{{\bm{\theta}}_{s}^{\perp}}\ |({\bm{\theta}}_{s}^{\perp})^{H}\bm{f}_{s}|^2\\
	&\ \mbox{s.t.} \ \  
({\bm{\theta}}_{s}^{\perp})^{H}\bm{f}_{p}=0,
\label{eq: ZF-zero}
\\
	&\quad \quad \ |{{\theta}}_{s,n}^{\perp}|=1, \forall n=1,\cdots,N. \label{eq: ZF-unit-modulus}
	\end{align}
\end{subequations}

The problem \textbf{P-ZF} is non-convex and thus difficult to be solved. To address this problem, we first examine its feasibility to see whether the constraint \eqref{eq: ZF-zero} can be satisfied under \eqref{eq: ZF-unit-modulus}, which is shown in the following Lemma.
\begin{lemma} \label{lemma: feasibility}
   (Feasibility of problem \textbf{P-ZF}) \textbf{P-ZF} is feasible if and only if the following condition is satisfied
    \begin{align} \label{eq: ZF-feasible-condition}
        2 \|\bm{f}_{p}\|_{\infty} \leq \|\bm{f}_{p}\|_{1},
    \end{align}
where $\|\bm{f}_{p}\|_{\infty}=\max \limits_{n=1,\cdots,N} \{|f_{p,n}|\} $, $\|\bm{f}_{p}\|_{1}=\sum_{n=1}^{N}|f_{p,n}|$, and $f_{p,n}$ denotes the $n$-th element of $\bm{f}_{p}$.
\end{lemma}
\begin{IEEEproof}
The proof is similar to \cite[Theorem 1 and Corollary 1]{zhang2016per}, which is thus omitted here for brevity.
\end{IEEEproof}

Lemma \ref{lemma: feasibility} shows that the condition in \eqref{eq: ZF-feasible-condition} can be simplified as $|f_{p,m}| \leq \sum_{n\neq m}^{N}|f_{p,n}|$, provided that $|f_{p,m}|=\max \limits_{n=1,\cdots,N} \{|f_{p,n}|\}$. This implies that the maximum of $\bm{f}_{p}$ should not be so dominant over the other elements of $\bm{f}_{p}$. Otherwise, even though the remaining elements achieve a coherent combination with the same phase by adjusting ${\bm{\theta}}_{s}^{\perp}$, the combined result is still smaller than the amplitude of the maximum $|f_{p,m}|$, which cannot achieve ZF for $({\bm{\theta}}_{s}^{\perp})^{H}\bm{f}_{p}=0$.

After checking the feasibility, we next use the alternating projection method~\cite{jiang2022interference} to obtain a suboptimal solution, whose idea is to alternatively project an initial solution onto the sets specified by the constraints. Specifically, \eqref{eq: ZF-zero} and \eqref{eq: ZF-unit-modulus} specify two sets of constraints, respectively, given by
\begin{align}
    \mathcal{S}_{1}&=\{{\bm{\theta}}_{s}^{\perp}:({\bm{\theta}}_{s}^{\perp})^{H}\bm{f}_{p}=0\} \\
    \mathcal{S}_{2}&=\{{\bm{\theta}}_{s}^{\perp}:|{{\theta}}_{s,n}^{\perp}|=1, \forall n=1,\cdots,N \}
\end{align}

Then, we choose the initial input ${{\bm{\theta}}_{s}^{\perp}}^{(0)}$ as $\bm{f}_{s}$, which could maximize the objective function $|({\bm{\theta}}_{s}^{\perp})^{H}\bm{f}_{s}|^2$.
With ${{\bm{\theta}}_{s}^{\perp}}^{(0)}$, we can alternately project it onto these two sets $\mathcal{S}_{1}$ and $\mathcal{S}_{2}$,
% Here, the initial input of ${{\bm{\theta}}_{s}^{\perp}}^{(0)}$ is chose as $\bm{f}_{s}^{H}$ which could maximize the objective function $|({\bm{\theta}}_{s}^{\perp})^{H}\bm{f}_{s}|^2$. 
% The basic idea of the alternating projection algorithm is to alternately project a solution onto these two sets $\mathcal{S}_{1}$ and $\mathcal{S}_{2}$ with an initial input ${{\bm{\theta}}_{s}^{\perp}}^{(0)}$,
which are given by the following closed-form expressions in the $q$-th iteration.
\begin{align}
{{\bm{\theta}}_{s}^{\perp}}^{(q)}&=\mathbf{Proj}_{\mathcal{S}_{1}}{{\bm{\theta}}_{s}^{\perp}}^{(q-1)} \nonumber \\
&=\bm{Z}_{p}{{\bm{\theta}}_{s}^{\perp}}^{(q-1)}={{\bm{\theta}}_{s}^{\perp}}^{(q-1)}-\frac{\bm{f}_{p}\bm{f}_{p}^{H}{{\bm{\theta}}_{s}^{\perp}}^{(q-1)}}{\|\bm{f}_{p}\|^2} ,
\label{eq: alternate-1}
\\
{{\bm{\theta}}_{s}^{\perp}}^{(q+1)}&=\mathbf{Proj}_{\mathcal{S}_{2}}{{\bm{\theta}}_{s}^{\perp}}^{(q)}={{\bm{\theta}}_{s}^{\perp}}^{(q)}/|{{\bm{\theta}}_{s}^{\perp}}^{(q)}|, \label{eq: alternate-2}
\end{align}
where $\bm{Z}_{p}$ denotes the projection operator that projects the signals into the space orthogonal to $\bm{f}_{p}$, given by
\begin{align}
    \bm{Z}_{p} = \bm{I}-\bm{f}_{p}(\bm{f}_{p}^{H}\bm{f}_{p})^{-1}\bm{f}_{p}^{H},
\end{align}

Through the above iterations in \eqref{eq: alternate-1} and \eqref{eq: alternate-2}, ${\bm{\theta}}_{s}^{\perp}$ is guaranteed to converge to a locally (globally) optimal point, for which the convergence analysis can be found in~\cite{jiang2022interference}.

% Obviously, without considering \eqref{eq: ZF-unit-modulus}, conventional ZF methods can be applied here to obtain a solution. ${\bm{\theta}}_{s}^{\perp}$ should be designed in the orthogonal complement of the subspace spanned by $\bm{f}_{p}$. Here, the projection operator which projects the signals into the space orthogonal to $\bm{f}_{p}$ is given by
% \begin{align}
%     \bm{Z}_{p} = \bm{I}-\bm{f}_{p}^{H}(\bm{f}_{p}\bm{f}_{p}^{H})^{-1}\bm{f}_{p},
% \end{align}

% With $\bm{Z}_{p}$, can be determined by 
% \begin{align}
% {\bm{\theta}}_{s}^{\perp}=\bm{f}_{s}\bm{Z}_{p}=\bm{f}_{s}-\frac{\bm{f}_{s}\bm{f}_{p}^{H}}{\|\bm{f}_{p}\|^2} \bm{f}_{p},
% \end{align}
% Alternate projection.

\subsubsection{Step 3 (Weighted MRT and ZF-based Beamforming Structure)}
With the obtained basis vectors $\bm{\theta}_{p}$ and ${\bm{\theta}}_{s}^{\perp}$, the proposed assistance-transmission beamforming structure is given by 
\begin{align} \label{eq: beamforming-structure}
\bm{\theta}=\bm{\theta}_{1}+c\bm{\theta}_{2}= \alpha \bm{\theta}_{p}+\beta c {\bm{\theta}}_{s}^{\perp},
\end{align}
where we let $\bm{\theta}_{1}=\alpha\bm{\theta}_{p}$, $\bm{\theta}_{2}=\beta\bm{\theta}_{s}^{\perp}$ with $\alpha$ and $\beta$ being two complex weighted parameters, and $c$ being the secondary signal transmitted at the RIS.
\subsection{Parameter Optimization of Proposed Assistance-Transmission Beamforming Structure}
Under the proposed structure \eqref{eq: beamforming-structure}, phase-shift
vector $\bm{\theta}$ is designed to lie in the subspace spanned by the two basis vectors $\bm{\theta}_{p}$ and $\bm{\theta}_{s}^{\perp}$. 
Besides, such a design enables us to optimize the two parameters $\alpha$ and $\beta$ only, which thus reduces the complexity significantly as compared to optimizing the $\bm{\theta}_{1}$ and $\bm{\theta}_{2}$ directly. To do so, we formulate a new problem \textbf{P2} with reference to \textbf{P1}, given by
\begin{subequations}
	\begin{align}
	\underline{\textbf{\text{P2:}}} \  
	&\max \limits_{\alpha,\beta}\ D_{\mathrm{su}}\\
	&\ \mbox{s.t.} \  \left| [\alpha \bm{\theta}_{p}+\beta c {\bm{\theta}}_{s}^{\perp}]_{n}\right|\leq 1, \forall c\in \mathcal{A}_{c}, n, \label{eq:P2-b} \\
	& \quad \quad  D_{\mathrm{pu}}\geq \eta, \label{eq:P2-c}\\
	& \quad \quad \eqref{eq:minimum-Euclidean-PU}, \eqref{eq:minimum-Euclidean-SU}. \nonumber
	\end{align}
\end{subequations}

Plugging \eqref{eq: beamforming-structure} into \eqref{eq:minimum-Euclidean-PU} and \eqref{eq:minimum-Euclidean-SU}, we can obtain the corresponding minimum Euclidean distances of the primary and secondary composite signals $D_{\mathrm{pu}}$ and $D_{\mathrm{su}}$, and further obtain the closed-form solutions of $\alpha$ and $\beta$, which will be presented subsequently.
\subsubsection{Optimization of $\alpha$}
First, we show the feasible region of $\textbf{P2}$, given by the following Proposition.
\begin{prop} \label{prop-optimization-alpha}
By solving the inequality of \eqref{eq:P2-c}, we can get the feasible region of $\alpha$, given by
 \begin{align} \label{eq:feasible-region-alpha}
    \alpha \geq \alpha_{0}
    = \sqrt{t^2+2\delta t+\delta}-t,
\end{align}
where $\delta$ is the assistance factor and $t$ is defined as the channel strength radio of the primary direct link to the primary reflecting link, given by
\begin{align} \label{eq-channel-strength-ratio}
    t\triangleq\frac{|h_p|}{\sum_{n=1}^{N}|f_{p,n}|}.
\end{align}
\end{prop}

Proposition \ref{prop-optimization-alpha} gives the minimum $\alpha$ that could satisfy the requirement of the primary system.
Obviously, considering the modulus constraint \eqref{eq:P2-b}, the optimal $\alpha^{\star}$ is obtained when the constraint \eqref{eq:feasible-region-alpha} is active since all the remaining energy can be assigned for the transmission beamforming vector $\beta\bm{\theta}_{s}^{\perp}$ to maximize the performance of the secondary system in this case.

Proposition \ref{prop-optimization-alpha} shows the following insights. The optimal $\alpha^{\star}$ depends on the assistance factor $\delta$, and the channel strength ratio $t$. On one hand, for a fixed $t$, when $\delta=0$, we have $\alpha^{\star}=0$. While $\delta=1$, it requires RIS to purely assist the primary system, and $\alpha^{\star}=1$ holds. On the other hand, for a fixed $\delta$, when $t=0$, i.e., the direct link is blocked, we have $\alpha^{\star}=\sqrt{\delta}$. While $t$ goes to infinity, we have $\alpha^{\star}=\delta$ according to \eqref{eq:feasible-region-alpha}
since $\sqrt{t^2+2\delta t+\delta}$ approaches $t+\delta$ via second-order Taylor series expansion $\sqrt{1+x}=1+\frac{1}{2}x-\frac{1}{8}x^2+\mathcal{O}(x^2)$.
\subsubsection{Optimization of $\beta$}
Given the optimized $\alpha$, we next optimize $\beta$, which is denoted by $\beta=|\beta|e^{\jmath\phi}$.
Considering the modulus constraints, the modulus of $\beta$ can be determined as $1-|\alpha^{\star}|$. Then, we need to determine the phase of $\beta$.
To do so, we first analyze the Euclidean distance of the secondary system, which is given by 
\begin{align} \label{eq: distance-secondary-expansion}
	D_{\mathrm{su}}\!=\!\begin{cases}
	|\bm{\theta}_{2}^{H}\bm{f}_{s}|^2|c_{m}-c_{k}|^2, \quad  \  s_{i}=s_{j},c_{m}\neq c_{k},\\
	|h_{s}\!+\!\bm{\theta}_{1}^{H}\bm{f}_{s}|^2|s_{i}-s_{j}|^2, s_{i}c_{m}\!=\!s_{j}c_{k},s_{i}\!\neq\! s_{j}, c_{m}\!\neq\! c_{k} \\
	\left|h_{s}\!+\!\bm{\theta}_{1}^{H}\bm{f}_{s}\!+\!\bm{\theta}_{2}^{H}\bm{f}_{s}\frac{s_{i}\!-\!s_{j}\frac{c_k}{c_m}}{s_{i}\!-\!s_{j}}c_m\right|^2\!|s_{i}\!-\!s_{j}|^2, \ \mathrm{otherwise}.
\end{cases} 
\end{align}

According to the optimization of $\alpha$, we have
$\bm{\theta}_{1}=\alpha^{\star}\bm{\theta}_{p}$, $\bm{\theta}_{2}=(1-|\alpha^{\star}|)e^{\jmath\phi}\bm{\theta}_{s}^{\perp}$. In this case, we see that the first and the second terms of $D_{\mathrm{su}}$ remain unchanged regardless of the phase of $\beta$. Therefore, the phase of $\beta$ can be determined by the third term of \eqref{eq: distance-secondary-expansion}, given by
\begin{align}
    \phi^{\star}=\arg \max\limits_{\phi\in [0,2\pi) } D_{\mathrm{su}}(\phi),
\end{align}
where $D_{\mathrm{su}}(\phi)=|h_{s}+\alpha^{\star}\bm{\theta}_{p}\bm{f}_{s}+(1-|\alpha^{\star}|)e^{\jmath\phi}\bm{\theta}_{s}^{\perp}\bm{f}_{s}\frac{s_{i}-s_{j}\frac{c_k}{c_m}}{s_{i}-s_{j}}|^2|s_{i}-s_{j}|^2$, $\forall c_{m}\neq c_{k}, s_{i}\neq s_{j}, s_{i}c_{m}\neq s_{j}c_{k}$.
\section{Simulation Results} \label{sec-simulation-results}
\vspace{-0.1cm}
In this section, simulation results are presented to validate the effectiveness of our proposed modulation scheme. Assume the primary and secondary signals adopt QPSK and BPSK as an example. 
The large-scale path loss of channels is modeled as $\mathrm{PL}(d,\xi)\!=\!10^{-3}d^{-\xi}$, where $d$ denotes the distance and $\xi$ denotes the path loss exponent. Besides, the small-scale fading of channels is assumed to follow Rayleigh distribution. The coordinates of PT, ST (RIS), PU, and SU are set to $x_{\mathrm{PT}}\!=\!(0,0)$,  $x_{\mathrm{ST}}\!=\!(0,30\mathrm{m})$, $x_{\mathrm{PU}}\!=\!(1000\mathrm{m},0)$, and $x_{\mathrm{SU}}\!=\!(990\mathrm{m},100\mathrm{m})$. Then, the distances of the channels $\bm{h}_{p}$, $\bm{h}_{s}$, $\bm{H}$, $\bm{g}_{p}$, and $\bm{g}_{s}$ can be calculated and whose path loss exponents are set to $2.9$, $2.8$, $2.1$, $2.3$, and $2.2$, respectively.
The maximum transmit power is set to $P_{t}\!=\!30$ dBm, and the noise power at the PU and SU are set to $\sigma_{p}^{2}\!=\!\sigma_{s}^{2}=-100$ dBm. For comparison, we use the conventional modulation scheme used in \cite{zhou2022cooperative,wang2021intelligent,hua2021novel} as a benchmark where the phase-shift matrix is set as $\bm{\Theta}(c)=c\bm{\Theta}_{2}$ with $\bm{\Theta}_1=\bm{0}$.
\begin{figure}[t]  
	\centering  
	% \captionsetup{font={scriptsize}}
	\setlength{\abovecaptionskip}{-0cm}
	\includegraphics[width=3in]{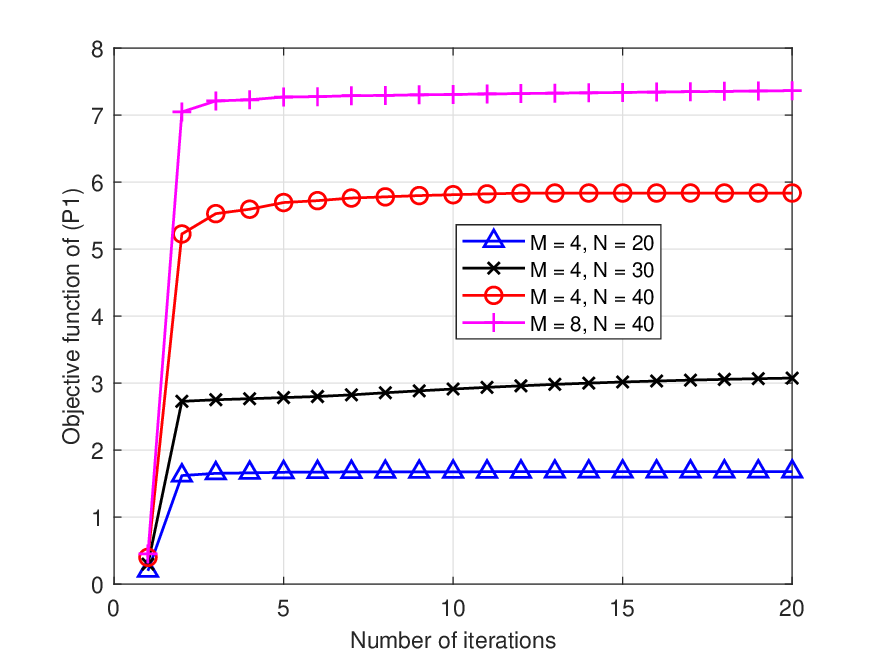} 
	\caption{Convergence behaviour for solving \textbf{P1}.} 
	\label{fig:convergence}  
\end{figure}

\begin{figure}[!t]
	\centering 
	\setlength{\abovecaptionskip}{-0.1cm}
% 	\setlength{\belowcaptionskip}{-0.05cm}
	% \captionsetup{font={scriptsize}}
	% \subfigcapskip = -10pt
	\subfigure[ $N=40$] {
		\label{fig:BER-tradeoff_N_40}
% 		\vspace{-0.5cm}
		\includegraphics[width=0.4\textwidth]{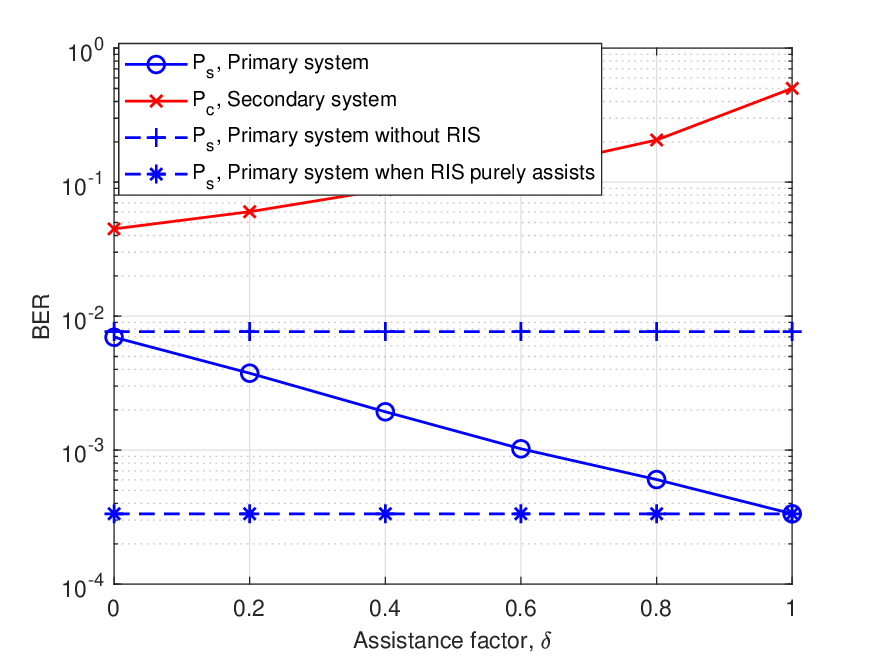}
	}
	\subfigure[$N=80$]{
		\label{fig:BER-tradeoff_N_80}
		\includegraphics[width=0.4\textwidth]{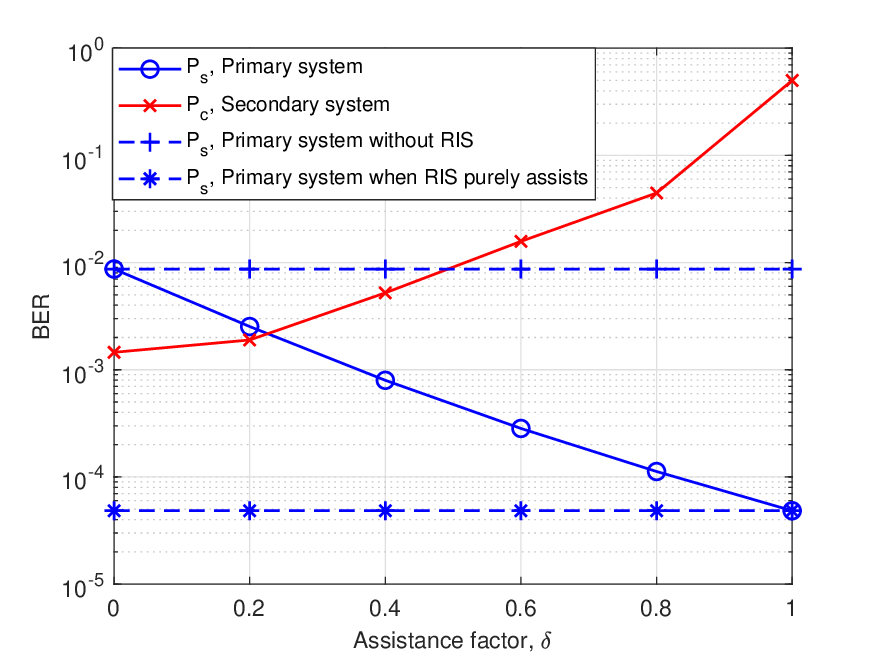}
	}
 %        \subfigure[$N=120$]{
	% 	\label{fig:BER-tradeoff_N_100}
	% 	\includegraphics[width=0.4\textwidth]{BER_assistance_factor_N_120.eps}
	% }
	\caption{BER performance of primary and secondary systems, $P_s$ and $P_c$, versus the assistance factor, $\delta$.}\label{fig: fig_tradeoff}
\end{figure}
\subsection{Validity of the Proposed Algorithms for \textbf{P1} }
First, we plot the convergence behavior of our proposed algorithm for solving \textbf{P1}. As shown in Fig. \ref{fig:convergence}, it is observed that our proposed algorithm could quickly converge within around $10$ iterations, which validates the effectiveness of the proposed algorithm. Moreover, with the increase of $M$ (number of transmit antennas) or $N$ (number of reflecting elements), the objective function of \textbf{P1}, i.e., the minimum Euclidean distance of the secondary composite signal, achieves a higher value due to the increased active/passive beamforming gain.

Next, in Fig. \ref{fig: fig_tradeoff}, we plot the BERs of primary and secondary systems, i.e., $P_{s}$ and $P_{c}$, versus the assistance factor $\delta$ by using the optimized assistance and transmission beamforming matrices from \textbf{P1}. From Fig. \ref{fig:BER-tradeoff_N_40}, with the increase of $\delta$, $P_s$ decreases while $P_c$ increases. This is because a higher assistance factor requires a larger $\bm{\Theta}_{1}$ to assist the primary system. In this case, due to the modulus constraint~\eqref{eq:P1-b}, there is only less space left for $\bm{\Theta}_{2}$ to transmit the secondary signal. Therefore, an interesting tradeoff between $P_{s}$ and $P_c$ is observed by adjusting the assistance factor $\delta$ and our proposed modulation scheme can strike a flexible balance between the BER of primary and secondary systems.
In practice, $\delta$ can be determined according to the performance requirement of the primary system. 
In addition, by varying $\delta$ from $0$ to $1$, $P_{s}$ is bounded between the BER performance under the two special RIS design schemes given by \eqref{eq: special-scheme-1} and \eqref{eq: special-scheme-2}.
When we look at Fig. \ref{fig:BER-tradeoff_N_80} that doubles $N$ as compared to Fig. \ref{fig:BER-tradeoff_N_40}, a similar tradeoff phenomenon can be observed. The difference is that the BER curves of the primary and secondary systems have an intersection point in Fig. \ref{fig:BER-tradeoff_N_80}. This is because increasing $N$ helps decrease the BER of the secondary system for the same $\delta$. Moreover, if we focus on the overall performance of $P_s$ and $P_c$, the intersection point would be the best to achieve the lowest BER for $P_s+P_c$.
% Also, similar observations can be made in Fig. \ref{fig:BER-tradeoff_N_100}.

\begin{figure}[t]  
	\centering  
	% \captionsetup{font={scriptsize}}
	% \setlength{\abovecaptionskip}{-0cm}
	% \setlength{\belowcaptionskip}{-0.5cm}
	\includegraphics[width=3in]{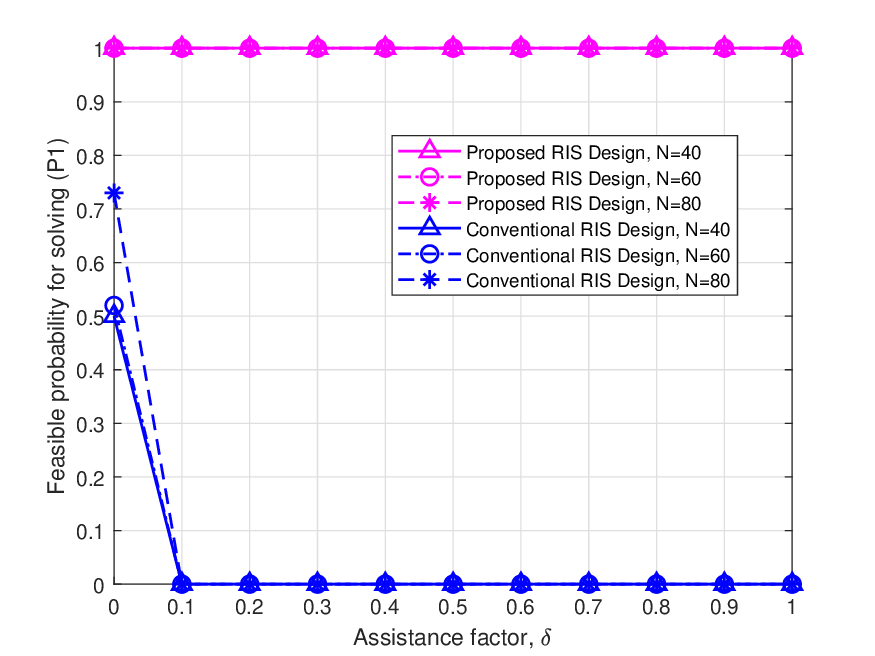} 
	\caption{Feasible probability for solving \textbf{P1} versus the assistance factor, $\delta$, under different RIS design schemes.} 
	\label{fig:distance-tradeoff}  
\end{figure}

Additionally, in Fig. \ref{fig:distance-tradeoff}, we compare the feasible probability for solving \textbf{P1} versus the assistance factor $\delta$ under different RIS design schemes. Here, the feasibility of \textbf{P1} lies in whether the performance requirement of the primary system, i.e., \eqref{eq:P1-c}, can be satisfied or not.
For the proposed RIS design scheme in \eqref{eq: phase-shifts-mapping}, no matter how large the $\delta$ is, one can observe that the problem \textbf{P1} can be successfully solved with a high probability close to $1$, which means that the proposed RIS design can satisfy higher performance requirement of the primary system.
However, for the conventional scheme, \textbf{P1} is feasible only when $\delta$ is very small. As for a large $\delta$, \textbf{P1} is infeasible and thus cannot satisfy the requirement of the primary system. This is because the conventional scheme, i.e., $\bm{\Theta}(c)=c\bm{\Theta}_{2}$, focuses on the information transmission for the secondary system, which could only provide limited assistance for the primary system. This motivates us to further investigate how much performance gain the conventional RIS design can offer to the primary system, which is studied in Fig. \ref{fig: BER-N-w/o-w-D}. 

\subsection{Impact of Number of Reflecting Elements on the BER Performance}
\begin{figure}[!t]
	\centering 
	% \setlength{\abovecaptionskip}{-0.1cm}
% 	\setlength{\belowcaptionskip}{-0.05cm}
	% \captionsetup{font={scriptsize}}
	% \subfigcapskip = -10pt
	\subfigure[ In the presence of the primary direct link] {
		\label{fig:BER-N} 
% 		\vspace{-0.5cm}
		\includegraphics[width=0.4\textwidth]{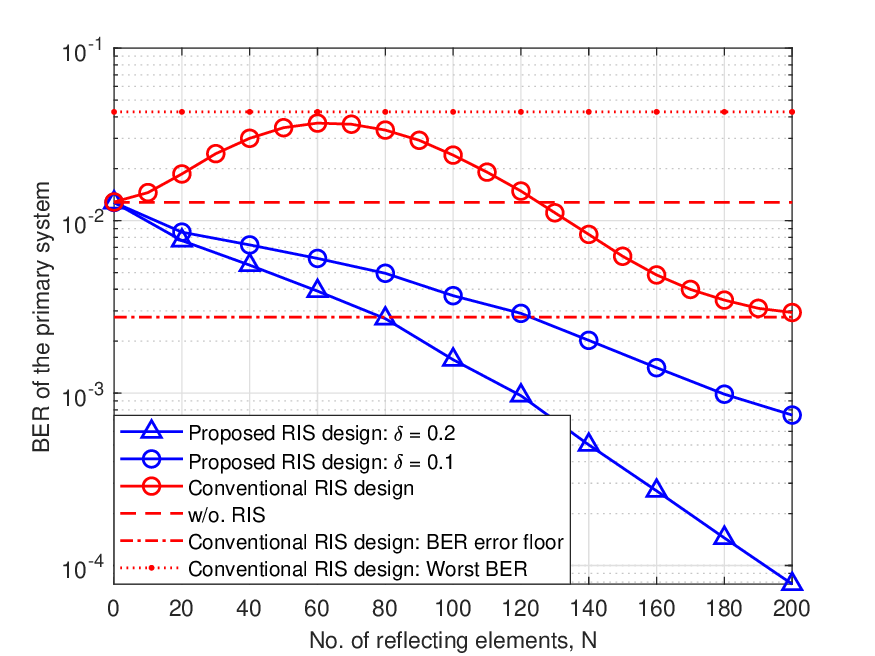}
	}
	\subfigure[In the absence of the primary direct link]{
		\label{fig:BER-wo-D-ambiguity} 
		\includegraphics[width=0.4\textwidth]{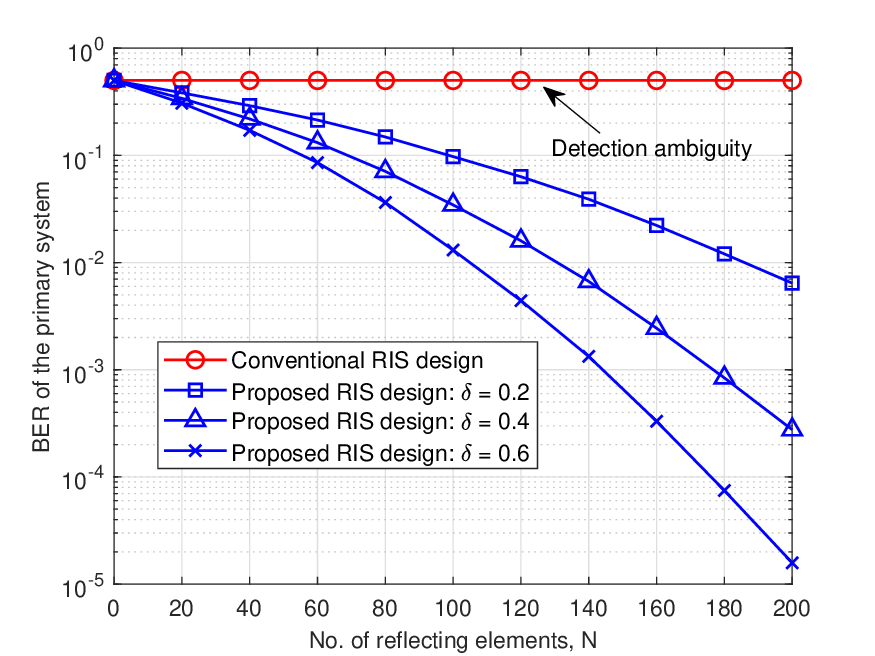}
	}
	\caption{BER performance of the primary system, $P_s$, versus the number of reflecting elements, $N$}\label{fig: BER-N-w/o-w-D}
\end{figure}

In Fig. \ref{fig: BER-N-w/o-w-D}, we study the impact of $N$ on the BER performance. To clearly show the BER error-floor problem of the conventional RIS design scheme, we focus on the performance of the primary system and fix the transmit beamforming vector as $\bm{w}=\sqrt{P_t}\frac{\bm{h}_{p}}{\Vert\bm{h}_{p}\Vert}$. Here, the phase shifts can be determined by maximizing the minimum Euclidean distance of the primary composite signal, which is given by $\theta_{n}=e^{\jmath(\frac{\pi}{4}+\angle (\bm{h}_{p}^{H}\bm{w})-\angle(\mathrm{diag}(\bm{g}_{p}^{H})\bm{H}\bm{w})_{n})}, \forall n=1,\cdots,N$, and the notation $(\bm{x})_{n}$ denotes the $n$-th element of $\bm{x}$.
From Fig. \ref{fig:BER-N}, we see that for the conventional RIS design scheme, with the increase of $N$, the BER of the primary system first increases and then decreases, and finally converges to a fixed value irrespective of $N$. Therefore, a BER error-floor phenomenon is observed, which validates the effectiveness of the analysis in Sec. \ref{sec:performance}. This is because under the conventional RIS design scheme, the minimum Euclidean distance of the primary composite signal becomes $d_3(N) = \sqrt{2}|\bm{h}_{p}^{H}\bm{w}+\bm{g}_{p}^{H}\bm{\Theta}_{2}\bm{H}\bm{w}|$ in \eqref{eq: distance-primary-conventional} when $0\leq N \leq 120$, which first decreases and then increases by increasing $N$. In this process, $d_{3}(N)$ achieves its minimum when $N=60$, which corresponds to the worst BER $\mathcal{Q}(\mu|\bm{h}_{p}^{H}\bm{w}|)$ where $\mu=\sqrt{\frac{P_t}{2\sigma_{p}^{2}}}$.
In cases $N$ is large enough, the minimum Euclidean distance of the primary composite signal becomes $d_2(N)\!=\!2|\bm{h}_p\bm{w}|$ for a large $N$. In this case, the BER converges to $\mathcal{Q}(2\mu|\bm{h}_p\bm{w}|)$, which implies that the BER of the primary transmission is limited by the strength of the equivalent direct link $\bm{h}_{p}\bm{w}$ regardless of $N$. Overall, the conventional RIS design scheme could only achieve marginal performance gain as compared to the case without RIS. By adopting our proposed modulation scheme, the BER error-floor problem can be readily addressed due to the introduction of the assistance beamforming matrix $\bm{\Theta}_{1}$. In this case, the equivalent direct link becomes $\bm{h}_{p}\bm{w}+\bm{g}_{p}^{H}\bm{\Theta}_{1}\bm{H}\bm{w}$, which can be enhanced with the increase of $N$. 

Furthermore, considering the direct link is blocked, we plot the BER of the primary system in Fig. \ref{fig:BER-wo-D-ambiguity}. It is observed that when the direct link is absent, the BER of the primary system is always $0.5$ regardless of the transmit power. This is due to the detection ambiguity when the direct link is blocked, which verifies the analysis in Sec. \ref{sec:performance}. Moreover, we can see for a large $\delta$, the BER performance of the primary system becomes better due to a higher performance requirement.
\subsection{Validity of BER Analysis and Proposed Low-Complexity Assistance-Transmission Beamforming Design}
\begin{figure}[t]  
	\centering  
	% \captionsetup{font={scriptsize}}
	\setlength{\abovecaptionskip}{-0cm}
	\includegraphics[width=3in]{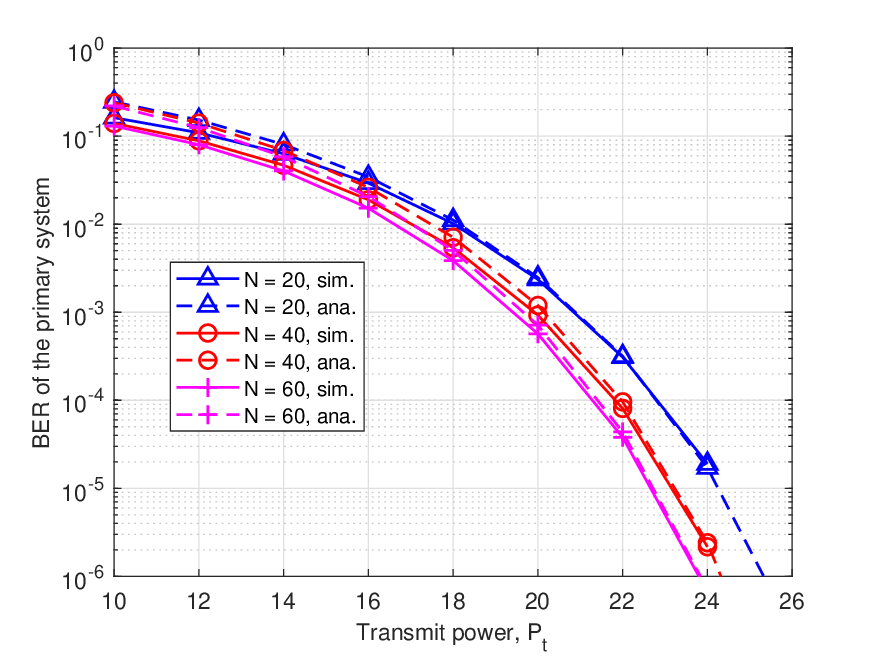}  
	\caption{BER performance of the primary system, $P_s$, versus the transmit power, with the assistance factor being $\delta=0.5$.}
 \vspace{-0.3cm}
	\label{fig: primary-BER-power}  
\end{figure}

\begin{figure}[t]  
	\centering  
	% \captionsetup{font={scriptsize}}
	\setlength{\abovecaptionskip}{-0cm}
	\includegraphics[width=3in]{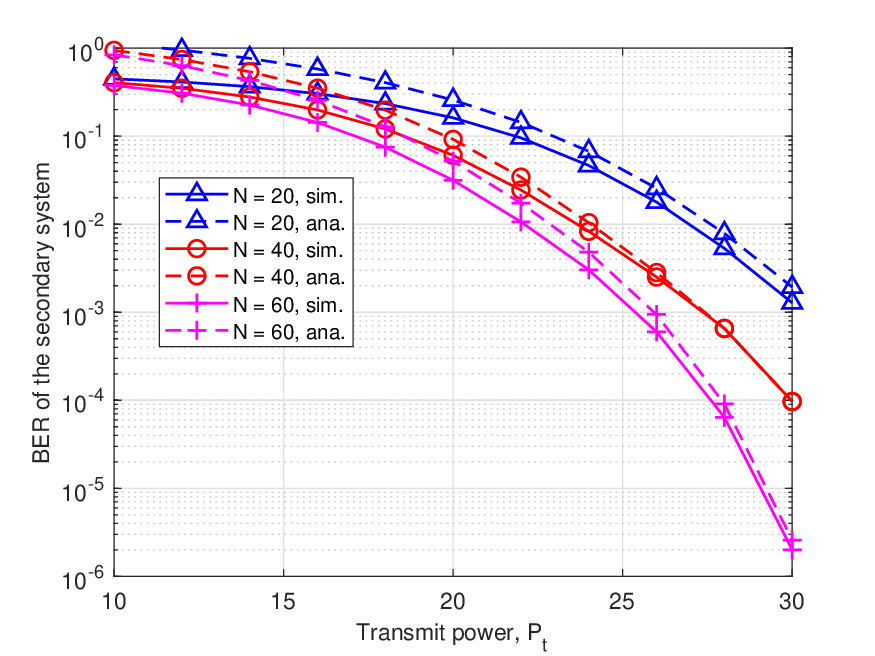}  
	\caption{BER performance of the secondary system, $P_c$, versus the transmit power, with the assistance factor being $\delta=0.5$.}
	\label{fig:BER-power}  
\end{figure}

In Fig. \ref{fig: primary-BER-power} and Fig. \ref{fig:BER-power}, we plot the BER of the secondary system versus the transmit power under different $N$. Note that the analytical results are plotted by using the union bound shown in \eqref{eq: Ps} and \eqref{eq: Pc}, while the simulation results are plotted by using the Monte Carlo method. It can be seen that the union bound is an upper bound of the exact BER, which performs close to the simulation results in the high SNR regime~\cite{proakis2001digital}. This validates the effectiveness of our BER analysis.
% Besides, we observe that the upper bound derived in \eqref{eq: Pc_upper} is indeed a looser form of the analytical results, which validates the accuracy of the BER approximation in Section. \ref{sec-distance}.
Moreover, it is observed that with the increase of $N$, a better BER performance for the primary and  secondary systems can be achieved due to the larger passive beamforming gain.

\begin{figure}[t]  
	\centering  
	% \captionsetup{font={scriptsize}}
	\setlength{\abovecaptionskip}{-0cm}
	\includegraphics[width=3in]{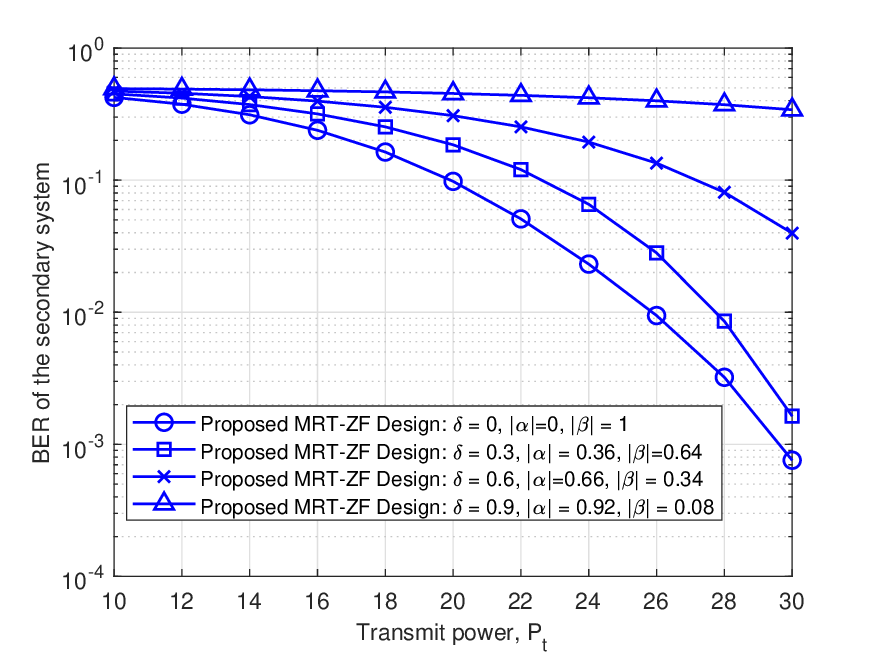}  
	\caption{BER performance of the secondary system, $P_c$, versus the transmit power, under the proposed MRT-ZF design with $N=40$.}
	\label{fig:MRT-ZF-BER-power}  
\end{figure}

\begin{figure}[t]  
	\centering  
	% \captionsetup{font={scriptsize}}
	\setlength{\abovecaptionskip}{-0cm}
	\includegraphics[width=3in]{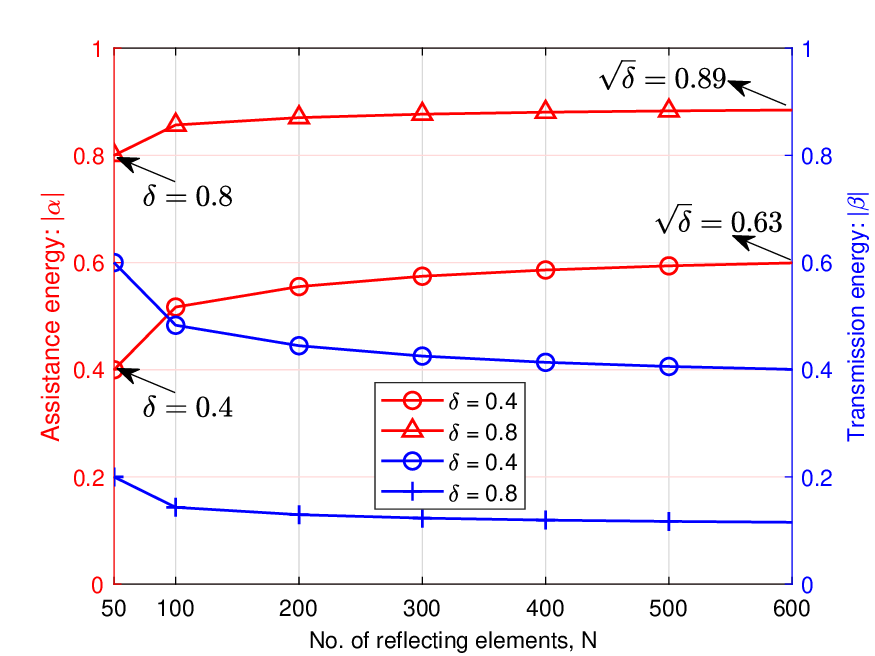}  
	\caption{Energy parameters of assistance and transmission beamforming matrices, $|\alpha|$ and $|\beta|$, versus the number of reflecting elements, $N$.}
	\label{fig:MRT-ZF-alpha-beta}  
\end{figure}

% \subsection{Validity of the Proposed Low-Complexity Assistance and Transmission Beamforming Design}
In Fig. \ref{fig:MRT-ZF-BER-power}, we evaluate the performance of the proposed MRT-ZF assistance-transmission beamforming structure and focus on the energy allocated to $\alpha$ and $\beta$. It is observed that with the increase of $\delta$, the BER of the secondary system decreases accordingly. For ease of understanding, we calculate the values of $\alpha$ and $\beta$ for a given $\delta$ based on \eqref{eq:feasible-region-alpha}. We can see that a higher $\delta$ requires a larger $|\alpha|$ to assist the primary system, while only a smaller $|\beta|$ is left for the secondary system due to the modulus constraints. This phenomenon validates the effectiveness of the proposed MRT-ZF assistance-transmission beamforming structure and clearly demonstrates the energy allocation for $\alpha$ and $\beta$ under different $\delta$, which provides useful guidance on the assistance-transmission beamforming design.

Finally, in Fig. \ref{fig:MRT-ZF-alpha-beta}, we plot the energy allocated for $\alpha$ and $\beta$ versus the number of reflecting elements $N$.
For the case of $\delta=0.4$,  with the increase of $N$, it is observed that $|\alpha|$ first increases from $\delta$ and then converges to a value of $\sqrt{\delta}$.
Conversely, $|\beta|$ first decreases from $1-\delta$ and then converges to a values of $1-\sqrt{\delta}$. This is because increasing $N$ will decrease the value of channel strength ratio $t$ defined in \eqref{eq-channel-strength-ratio}. When $N$ is very large, $t$ goes to zero. In this case, $|\alpha|$ converges to $\sqrt{\delta}$ based on \eqref{eq-channel-strength-ratio}. Moreover, a similar phenomenon can be observed in the case of $\delta=0.8$. On the other hand, for a fixed number of reflecting elements $N$, it can be seen that the allocated energy for $|\alpha|$ under $\delta=0.8$ is larger than that under $\delta=0.4$, which is expected since a higher $\delta$ needs a larger $\alpha$ for assisting the primary system.
% analysis \eqref{eq: Pc} is the union bound of BER, which is tight in high SNR regime.
% where the analytical results and the upper bound are obtained using \eqref{eq: Pc} and \eqref{eq: Pc_upper}. Note that the analysis \eqref{eq: Pc} is the union bound of BER, which is tight in high SNR regime.
% We see that the analytical results approaches the simulation results in high SNR regime, thereby validating the effectiveness of the performance analysis in Section. \ref{sec: design criterion}. 
% We see that the BER approximation in \eqref{eq: Pc_upper} serves an upper bound of $P_c$, which means that optimizing the minimum Eucldiean distance is effective for BER minimization.

% due to the larger passive beamforming gain brought by RIS.

\vspace{-0.1cm}
\section{Conclusion}\label{sec-conslusion}
\vspace{-0.1cm}
In this paper, we have proposed a novel over-the-air modulation scheme for RIS-assisted SR, which divides the RIS phase-shift matrix into two parts, where one is the assistance beamforming matrix used for assisting the primary system and the other is the transmission beamforming matrix used for transmitting the secondary signal. To optimize the assistance and transmission beamforming matrices, we have introduced an assistance factor and then formulated a problem to minimize the BER of the secondary system while guaranteeing that the BER performance of the primary system is higher than the performance requirement controlled by the assistance factor. By solving the problem, simulation results have shown that our proposed RIS design scheme could strike a balance between the performance of the primary and secondary systems under different assistance factors.
\vspace{-0.15cm}
\begin{appendices} 
\section{} \label{appendix-A}
According to \cite{proakis2001digital}, the BER of decoding $s$ at the PU can upper-bounded by
\begin{align}
    P_{s} &= 
       \sum_{i=1}^{|\mathcal{A}_{s}|} \sum_{m=1}^{|\mathcal{A}_{c}|}\sum_{j=1,j\neq i}^{|\mathcal{A}_{s}|} \sum_{k=1}^{|\mathcal{A}_{c}|} \mathcal{Q}\left(\sqrt{\frac{P_{t}d_{p}((s_{i},c_{m}),(s_{j},c_{k}))}{2\sigma_{p}^{2}}}\right)\nonumber\\  &\quad \quad \quad \quad \quad \quad \quad \quad \quad \times  \frac{1}{|\mathcal{A}_{s}||\mathcal{A}_{c}|}\frac{e(s_{i}\rightarrow s_{j})}{\log_{2}(|\mathcal{A}_{s}|)} \nonumber\\
    &\overset{(a)}{\leq}  (|\mathcal{A}_{s}|-1)|\mathcal{A}_{c}|\mathcal{Q}\left(\sqrt{\frac{P_{t}D_{p}}{2\sigma_{p}^{2}}}\right) \nonumber\\ 
    &\overset{(b)}{\leq} \frac{1}{2}(|\mathcal{A}_{s}|-1)|\mathcal{A}_{c}| e^{-\frac{P_{t}D_{p}}{4\sigma_{p}^{2}}},
    \label{eq: Ps_upper}
\end{align}
where (a) holds since $\mathcal{Q}$-function is a decreasing function and we scale all the Hamming distances $e(s_{i}\!\rightarrow \!s_{j})$ to its maximum, i.e., $\log_{2}(\mathcal{A}_{s})$; (b) follows from the fact $\mathcal{Q}(t)\leq\frac{1}{2}e^{-\frac{1}{2}t^2}$; $D_{p}$ denotes the minimum Euclidean distance of the primary composite signal, which is defined in \eqref{eq:minimum-Euclidean-PU}.

Similarly, the upper bound of $P_{c}$ at the SU is derived as
\begin{align}
    P_{c}&\leq \frac{1}{2}|\mathcal{A}_{s}|(|\mathcal{A}_{c}|-1) e^{-\frac{P_{t}D_{s}}{4\sigma_{s}^{2}}}. \label{eq: Pc_upper}
\end{align}
In \eqref{eq: Pc_upper}, $D_{s}$ denotes the minimum Euclidean distance of the secondary composite signal, which is defined in \eqref{eq:minimum-Euclidean-SU}.
\end{appendices}
% strike a flexible balance between the performance of the primary and secondary transmissions by optimizing the two sub-phase-shift matrices. 

% In this paper, we have studied the RIS design for SR from the perspective of assistance-transmission tradeoff. Specifically, a novel RIS partitioning design has been proposed to realize such a tradeoff. Then,  the assistance-transmission tradeoff problem of optimizing the surface partitioning strategy has been formulated to minimize the BER of the composite signal, so as to strike a balance between the BER performance of the primary and secondary signals. By solving this problem, we have shown that there exists the optimal surface partitioning strategy, which provides the best tradeoff. Finally, simulation results have shown that the proposed RIS design outperforms the conventional counterpart significantly.
\bibliographystyle{IEEEtran}
\bibliography{refFile}

\begin{thebibliography}{10}
\providecommand{\url}[1]{#1}
\csname url@samestyle\endcsname
\providecommand{\newblock}{\relax}
\providecommand{\bibinfo}[2]{#2}
\providecommand{\BIBentrySTDinterwordspacing}{\spaceskip=0pt\relax}
\providecommand{\BIBentryALTinterwordstretchfactor}{4}
\providecommand{\BIBentryALTinterwordspacing}{\spaceskip=\fontdimen2\font plus
\BIBentryALTinterwordstretchfactor\fontdimen3\font minus \fontdimen4\font\relax}
\providecommand{\BIBforeignlanguage}[2]{{%
\expandafter\ifx\csname l@#1\endcsname\relax
\typeout{** WARNING: IEEEtran.bst: No hyphenation pattern has been}%
\typeout{** loaded for the language `#1'. Using the pattern for}%
\typeout{** the default language instead.}%
\else
\language=\csname l@#1\endcsname
\fi
#2}}
\providecommand{\BIBdecl}{\relax}
\BIBdecl

\bibitem{zhou2023}
H.~{Zhou} and Y.-C. {Liang}, ``{RIS} design for symbiotic radio: A mutualistic spectrum sharing perspective,'' in \emph{Proc. IEEE Global Commun. Conf. (Globecom)}.\hskip 1em plus 0.5em minus 0.4em\relax Kuala Lumpur, Malaysia: IEEE, 2023, pp. 116--121.

\bibitem{nguyen20216g}
D.~C. Nguyen, M.~Ding, P.~N. Pathirana, A.~Seneviratne, J.~Li, D.~Niyato, O.~Dobre, and H.~V. Poor, ``{6G Internet of Things}: A comprehensive survey,'' \emph{IEEE Internet Things J.}, vol.~9, no.~1, pp. 359--383, 2021.

\bibitem{dang2020should}
S.~Dang, O.~Amin, B.~Shihada, and M.-S. Alouini, ``What should {6G} be?'' \emph{Nature Electronics}, vol.~3, no.~1, pp. 20--29, 2020.

\bibitem{long2019symbiotic}
R.~Long, Y.-C. Liang, H.~Guo, G.~Yang, and R.~Zhang, ``Symbiotic radio: A new communication paradigm for passive internet of things,'' \emph{IEEE Internet Things J.}, vol.~7, no.~2, pp. 1350--1363, 2019.

\bibitem{liang2020symbiotic}
Y.-C. Liang, Q.~Zhang, E.~G. Larsson, and G.~Y. Li, ``Symbiotic radio: Cognitive backscattering communications for future wireless networks,'' \emph{IEEE Trans. Cogn. Commun. Netw.}, vol.~6, no.~4, pp. 1242--1255, 2020.

\bibitem{wang2024multi}
J.~Wang, Y.-C. Liang, and S.~Sun, ``Multi-user multi-{IoT}-device symbiotic radio: A novel massive access scheme for cellular {IoT},'' \emph{IEEE Trans. Wireless Commun.}, 2024, early access, 10.1109/TWC.2024.3385530.

\bibitem{dai2022rate}
Z.~Dai, R.~Li, J.~Xu, Y.~Zeng, and S.~Jin, ``Rate-region characterization and channel estimation for cell-free symbiotic radio communications,'' \emph{IEEE Trans. Commun.}, vol.~71, no.~2, pp. 674--687, 2022.

\bibitem{wu2019towards}
Q.~Wu and R.~Zhang, ``Towards smart and reconfigurable environment: Intelligent reflecting surface aided wireless network,'' \emph{IEEE Commun. Maga.}, vol.~58, no.~1, pp. 106--112, 2019.

\bibitem{huang2019reconfigurable}
C.~Huang, A.~Zappone, G.~C. Alexandropoulos, M.~Debbah, and C.~Yuen, ``Reconfigurable intelligent surfaces for energy efficiency in wireless communication,'' \emph{IEEE Trans. Wireless Commun.}, vol.~18, no.~8, pp. 4157--4170, 2019.

\bibitem{liang2019large}
Y.-C. Liang, R.~Long, Q.~Zhang, J.~Chen, H.~V. Cheng, and H.~Guo, ``Large intelligent surface/antennas ({LISA}): Making reflective radios smart,'' \emph{J. Commun. Info. Netw.}, vol.~4, no.~2, pp. 40--50, 2019.

\bibitem{lei2021reconfigurable}
X.~Lei, M.~Wu, F.~Zhou, X.~Tang, R.~Q. Hu, and P.~Fan, ``Reconfigurable intelligent surface-based symbiotic radio for 6{G}: Design, challenges, and opportunities,'' \emph{IEEE Wireless Commun.}, vol.~28, no.~5, pp. 210--216, 2021.

\bibitem{liang2022backscatter}
Y.-C. Liang, Q.~Zhang, J.~Wang, R.~Long, H.~Zhou, and G.~Yang, ``Backscatter communication assisted by reconfigurable intelligent surfaces,'' \emph{Proc. IEEE}, vol. 110, no.~9, pp. 1339--1357, 2022.

\bibitem{zhang2024channel}
Q.~Zhang, H.~Zhou, Y.-C. Liang, W.~Zhang, and H.~V. Poor, ``Channel capacity of {RIS}-assisted symbiotic radios with imperfect knowledge of channels,'' \emph{IEEE Trans. Cogn. Commun. Netw.}, 2024, early access, 10.1109/TCCN.2024.3379406.

\bibitem{chen2023transmission}
H.~Chen, R.~Long, and Y.-C. Liang, ``Transmission protocol and beamforming design for {RIS}-assisted symbiotic radio over {OFDM} carriers,'' in \emph{Proc. IEEE Global Commun. Conf. (Globecom)}.\hskip 1em plus 0.5em minus 0.4em\relax Kuala Lumpur, Malaysia: IEEE, 2023, pp. 3258--3263.

\bibitem{yan2020passive}
W.~Yan, X.~Yuan, Z.-Q. He, and X.~Kuai, ``Passive beamforming and information transfer design for reconfigurable intelligent surfaces aided multiuser {MIMO} systems,'' \emph{IEEE J. Sel. Areas Commun.}, vol.~38, no.~8, pp. 1793--1808, 2020.

\bibitem{hua2021novel}
M.~Hua, Q.~Wu, L.~Yang, R.~Schober, and H.~V. Poor, ``A novel wireless communication paradigm for intelligent reflecting surface based symbiotic radio systems,'' \emph{IEEE Trans. Signal Process.}, vol.~70, pp. 550--565, 2021.

\bibitem{hua2021uav}
M.~Hua, L.~Yang, Q.~Wu, C.~Pan, C.~Li, and A.~L. Swindlehurst, ``{UAV}-assisted intelligent reflecting surface symbiotic radio system,'' \emph{IEEE Trans. Wireless Commun.}, vol.~20, no.~9, pp. 5769--5785, 2021.

\bibitem{zhang2021reconfigurable}
Q.~Zhang, Y.-C. Liang, and H.~V. Poor, ``Reconfigurable intelligent surface assisted {MIMO} symbiotic radio networks,'' \emph{IEEE Trans. Commun.}, vol.~69, no.~7, pp. 4832--4846, 2021.

\bibitem{zhou2022cooperative}
H.~Zhou, X.~Kang, Y.-C. Liang, S.~Sun, and X.~Shen, ``Cooperative beamforming for reconfigurable intelligent surface-assisted symbiotic radios,'' \emph{IEEE Trans. Veh. Technol.}, vol.~71, no.~11, pp. 11\,677--11\,692, 2022.

\bibitem{guo2020reflecting}
S.~Guo, S.~Lv, H.~Zhang, J.~Ye, and P.~Zhang, ``Reflecting modulation,'' \emph{IEEE J. Sel. Areas Commun.}, vol.~38, no.~11, pp. 2548--2561, 2020.

\bibitem{wu2021reconfigurable}
M.~Wu, X.~Lei, X.~Zhou, Y.~Xiao, X.~Tang, and R.~Q. Hu, ``Reconfigurable intelligent surface assisted spatial modulation for symbiotic radio,'' \emph{IEEE Trans. Veh. Technol.}, vol.~70, no.~12, pp. 12\,918--12\,931, 2021.

\bibitem{li2022ris}
Q.~Li, S.~Bai, J.~Li, Z.~Hu, and J.~Wang, ``{RIS}-assisted joint active and passive transmission with distributed reception,'' \emph{IEEE Trans. Veh. Technol.}, vol.~72, no.~5, pp. 6805--6809, 2023.

\bibitem{lin2020reconfigurable}
S.~Lin, B.~Zheng, G.~C. Alexandropoulos, M.~Wen, M.~Di~Renzo, and F.~Chen, ``Reconfigurable intelligent surfaces with reflection pattern modulation: Beamforming design and performance analysis,'' \emph{IEEE Trans. Wireless Commun.}, vol.~20, no.~2, pp. 741--754, 2020.

\bibitem{lin2021reconfigurable}
S.~Lin, F.~Chen, M.~Wen, Y.~Feng, and M.~Di~Renzo, ``Reconfigurable intelligent surface-aided quadrature reflection modulation for simultaneous passive beamforming and information transfer,'' \emph{IEEE Trans. Wireless Commun.}, vol.~21, no.~3, pp. 1469--1481, 2021.

\bibitem{li2022reconfigurable}
Q.~Li, M.~Wen, L.~Xu, and K.~Li, ``Reconfigurable intelligent surface-aided number modulation for symbiotic active/passive transmission,'' \emph{IEEE Internet Things J.}, vol.~10, no.~22, pp. 19\,356--19\,367, 2023.

\bibitem{wang2021intelligent}
C.~Wang, Z.~Li, T.-X. Zheng, D.~W.~K. Ng, and N.~Al-Dhahir, ``Intelligent reflecting surface-aided secure broadcasting in millimeter wave symbiotic radio networks,'' \emph{IEEE Trans. Veh. Technol.}, vol.~70, no.~10, pp. 11\,050--11\,055, 2021.

\bibitem{hu2020reconfigurable}
J.~Hu, Y.-C. Liang, and Y.~Pei, ``Reconfigurable intelligent surface enhanced multi-user {MISO} symbiotic radio system,'' \emph{IEEE Trans. Commun.}, vol.~69, no.~4, pp. 2359--2371, 2020.

\bibitem{wu2023ris}
M.~Wu, X.~Lei, X.~Zhou, X.~Tang, and O.~A. Dobre, ``{RIS}-assisted energy-and spectrum-efficient symbiotic transmission in {NOMA} systems,'' \emph{IEEE Trans. Commun.}, vol.~70, no.~12, pp. 12\,918--12\,931, 2021.

\bibitem{ye2020joint}
J.~Ye, S.~Guo, and M.-S. Alouini, ``Joint reflecting and precoding designs for {SER} minimization in reconfigurable intelligent surfaces assisted {MIMO} systems,'' \emph{IEEE Trans. Wireless Commun.}, vol.~19, no.~8, pp. 5561--5574, 2020.

\bibitem{lo1999maximum}
T.~K. Lo, ``Maximum ratio transmission,'' in \emph{IEEE Int. Conf. Commun. (Cat. No. 99CH36311)}, vol.~2.\hskip 1em plus 0.5em minus 0.4em\relax IEEE, 1999, pp. 1310--1314.

\bibitem{song2013prior}
S.~Song, M.~O. Hasna, and K.~B. Letaief, ``Prior zero forcing for cognitive relaying,'' \emph{IEEE Trans. Wireless Commun.}, vol.~12, no.~2, pp. 938--947, 2013.

\bibitem{zhou2023modulation}
H.~Zhou, B.~Cai, Q.~Zhang, R.~Long, Y.~Pei, and Y.-C. Liang, ``Modulation design and optimization for {RIS}-assisted symbiotic radios,'' \emph{arXiv preprint arXiv:2311.01167}, 2023.

\bibitem{zhou2023assistance}
H.~Zhou, Q.~Zhang, Y.-C. Liang, and Y.~Pei, ``Assistance-transmission tradeoff for {RIS}-assisted symbiotic radios,'' \emph{IEEE Trans. Wireless Commun.}, 2023, early access, doi:10.1109/TWC.2023.3335111.

\bibitem{proakis2001digital}
J.~G. Proakis, \emph{Digital communications}.\hskip 1em plus 0.5em minus 0.4em\relax Fourth Edition. New York: McGrawHill, Inc, 2001.

\bibitem{wu2019intelligent}
Q.~Wu and R.~Zhang, ``Intelligent reflecting surface enhanced wireless network via joint active and passive beamforming,'' \emph{IEEE Trans. Wireless Commun.}, vol.~18, no.~11, pp. 5394--5409, 2019.

\bibitem{zuo2022joint}
J.~Zuo, Y.~Liu, Z.~Ding, L.~Song, and H.~V. Poor, ``Joint design for simultaneously transmitting and reflecting ({STAR}) {RIS} assisted {NOMA} systems,'' \emph{IEEE Trans. Wireless Commun.}, vol.~22, no.~1, pp. 611--626, 2022.

\bibitem{grant2014cvx}
\BIBentryALTinterwordspacing
M.~Grant and S.~Boyd, ``{CVX}: Matlab software for disciplined convex programming, version 2.2,,'' Jan. 2020. [Online]. Available: \url{http:// cvxr.com/cvx.}
\BIBentrySTDinterwordspacing

\bibitem{zhang2016per}
J.~Zhang, Y.~Huang, J.~Wang, B.~Ottersten, and L.~Yang, ``Per-antenna constant envelope precoding and antenna subset selection: A geometric approach,'' \emph{IEEE Trans. Signal Process.}, vol.~64, no.~23, pp. 6089--6104, 2016.

\bibitem{jiang2022interference}
T.~Jiang and W.~Yu, ``Interference nulling using reconfigurable intelligent surface,'' \emph{IEEE J. Sel. Areas Commun.}, vol.~40, no.~5, pp. 1392--1406, 2022.

\end{thebibliography}
\end{document}